\documentclass[12pt,preprint]{aastex}
\slugcomment{Accepted and scheduled for publication  
in {\it The Astrophysical Journal},  for the ApJ December 10, 2009, v 707, 1 issue}  
\def\lax {\ifmmode{_<\atop^{\sim}}\else{${_<\atop^{\sim}}$}\fi}  
\def\gax {\ifmmode{_>\atop^{\sim}}\else{${_>\atop^{\sim}}$}\fi}  
\def\gtorder{\mathrel{\raise.3ex\hbox{$>$}\mkern-14mu
             \lower0.6ex\hbox{$\sim$}}}
\def\etal { et al. }

\begin{document}

\title{Discovery of  photon index saturation  in the black hole binary GRS~1915+105}
%\title{Spectral and Variability Characteristics of Black Hole binary GRS~1915+105}

\author{Lev Titarchuk\altaffilmark{1} and Elena Seifina\altaffilmark{2}}

\altaffiltext{1}{Dipartimento di Fisica, Universit\`a di Ferrara, Via 
Saragat 1, I-44100 Ferrara, Italy, email:titarchuk@fe.infn.it; ICRANET Piazzale d. Repubblica 10-12
65122 Pescara,  Italy; George Mason University Fairfax, VA 22030; and US Naval Research
Laboratory, Code 7655, Washington, DC 20375; email:Lev.Titarchuk@nrl.navy.mil;  
Goddard Space Flight Center, NASA,  code 663, Greenbelt  
MD 20771, USA} 

\altaffiltext{2}{Moscow State University/Sternberg Astronomical Institute, Universitetsky 
Prospect 13, Moscow, 119992, Russia; seif@sai.msu.ru}

%\altaffiltext{2}{Goddard Space Flight Center, NASA, 
%Astrophysics Science Division, code 662, Greenbelt MD 20771}

\begin{abstract}
%We investigate the evolution of spectral and timing properties in a number of 
%Galactic black hole (BH) sources during spectral transitions.
% in order
%to study the implications of scaling of a correlation between  photon index of Comptonized component 
%and the  centroid  of  low frequency quasi-periodic  oscillations (QPO) in the X-ray energy spectrum. 
%Using recently improved 
%RXTE response we analyze energy spectra of these BH sources in a broad  energy range from 3 keV 
%to 60 keV. 
%We also  analyze  the evolution  of Fourier Power Density Spectra along with the energy spectral evolution. Particularly we are  interested in the behavior  
%of low-frequency quasi-periodic  oscillations (QPOs) vs photon index (spectral state).
We present a study of the correlations between spectral, timing properties and mass accretion rate observed in X-rays from the
Galactic Black Hole (BH) binary GRS 1915+105 during the transition
between hard and soft states. 
We analyze 
%almost 
all transition  episodes 
from this source  observed with { Rossi X-ray Timing Explorer}
({\it RXTE}), 
{coordinated with Ryle Radio Telescope (RT) observations.
We show that  broad-band energy spectra of GRS~1915+105 during all these spectral states can be adequately presented   by  two  Bulk Motion Comptonization (BMC) components: a  hard component  (BMC1,  photon index    
  $\Gamma_1=1.7-3.0$)  with turnover at  high energies and soft thermal component  (BMC2, $\Gamma_2=2.7-4.2$) with characteristic 
color temperature $\le$~1 keV, and the redskewed iron line (LAOR) component. 
} 
%Specifically, we  demonstrate that during the  powerful X-ray outburst, %\textbf
%{followed} by a strong radio-flare, the power spectrum of X-ray signal is featureless.
%Before and after this outburst event, the power spectrum shows a typical  white-red noise shape along with the strong quasiperiodic oscillation (QPO) features in it. 
%We modify our scaling 
We also present observable correlations between
% photon index versus quasi-periodic oscillations along with the correlation between 
the index and the normalization of the  disk ``seed'' component.
% (mass accretion rate).
%These correlations one contains information on the relative BH mass with respect to that of other BH binary  the latter is expected to depend both on  the BH mass and the source distance for GRS 1915+105. This provides us with additional  observational constrain on mass and allows independent 
% evaluation  of the distance to the source.  
The use of ``seed'' disk  normalization, which is presumably proportional to mass accretion rate in the disk, is crucial to establish the index saturation effect during the transition to the soft state.
%as QPOs tend to disappear during the transition and does not usually allow to establish the saturation. 
We discovered the  photon index saturation of the soft and hard spectral components 
at values of $\lax$ 4.2 and 3  respectively.
%  in the Black Hole binary GRS~1915+105}
%\title
We   present a physical model which explains the index-seed photon normalization correlations.
We argue that the index saturation effect of the hard component (BMC1)  is due to the soft photon Comptonization in the converging inflow close to BH and that of soft component is due to matter accumulation in the transition layer when mass accretion rate increases.
%and therefore we propose it as a signature of the converging flow (black hole). 
Furthermore we demonstrate  a  strong correlation between equivalent width of the  iron line and radio flux in  GRS~1915+105. 
In addition to our spectral model components we  also  find  a strong feature of ``blackbody-like'' bump which  color temperature is about 4.5 keV  in  eight  observations of the intermediate and soft states. 
%that 
% the detected  photon spectrum is described by the sum of 
%two BMC, Laor and ``blackbody'' components with $kT_{bb}=$4.1-4.7 keV. 
We discuss  a possible origin of this   ``blackbody-like'' emission.
% in frame of ...(our) model.

%We use GRO J1655-40
%as the primary reference source as this system has the best determined
%mass, distance and inclination among Galactic BH sources.
%We apply scaling technique to determine BH masses and distances for
%Cygnus X-1,GX 339-4, 4U 1543-47, XTE J1550-564, XTE J1650-500, H 1743-322 and
%XTE J1859-226.
%In most cases the results of scaling are in good agreement
%with independent measurements, which
%provides a verification for our scaling method. 
%In the theoretical part of the Paper 
% We show the arguments for  the index saturation effect as an observational
%signature of the converging flow near BH horizon.   

\end{abstract}


\section{Introduction}

%Determination of masses of stellar mass black holes (BHs) is 
%one of the most important tasks in modern astronomy. Knowledge of
%mass distribution of the Galactic BHs can provide important clues
%about stellar evolution. It also can constrain the maximum mass
%of a neutron star with can form during a supernova explosion,
%which is believed to be about 3.2 solar masses \citep{rr74} but have not been 
%observationally verified. 
The study of the characteristic changes in spectral 
and variability properties 
of X-ray binaries is proved to be a valuable source of information on the physics 
governing the accretion processes
and on the fundamental parameters of black holes (BHs).
% A number of proposals
%were put forward to utilize X-ray spectral and timing data 
%to infer accreting BH masses. A temperature of the thermal component 
%in energy spectra was proposed as an indicator of BH mass \citet{soria08}.
%In this Paper we explore the implications
%of scaling laws to masses and distances of Galactic BHs systems
%using their X-ray spectral and timing properties.
%We concentrate our efforts on the study of the
%correlation between the spectral index and 
%characteristic sub-second 
%variability frequencies  and
%the  accretion disk luminosity. 
%More fundamentally, 
%We argue that the shape of the correlation
%pattern can contain the direct BH signature. 
%Namely, we show both observationally and theoretically 
%that the index saturates with mass accretion rate  which is a signature of a converging 
%flow. This index saturation effect  can exist only in  BH sources.
%Also this  correlation 
%pattern carries the most direct
%information on the BH mass and the distance to the source (see \cite{ST09}, hereafter ST09).

%frequency during state transitions in several Galactic BH transients
%and their application to BH mass determination.

The simultaneous study of the spectral and timing evolution of a BH source during 
a state transition has been a subject of many investigations 
[see references in a review by  \cite{rm}].
 \cite{fb04}, hereafter FB04,  introduced a classification  of  the spectral states in GRS 1915 +105 and studied  the spectral state evolution.   Using X-ray colors (hardness ratio) they introduced three spectral states.
State A: in which the strong blackbody-like (BB)  component of color temperature $\gax 1$ keV dominates in the overall spectrum and little time variability  is detected. State B: similar to state A but substantial red-noise variability  on scales $>1$ s occurs in this state.
State C: the spectra are harder than those in states A and B. Photon indices of the power-law components vary from 1.8 to 2.5.  White-red noise (WRN) variability  on scales $>1$ s  takes place in this state. 

 Furthermore FB04 discussed the connection between states A, B, C observed in GRS 1915 with
 the three ``canonical'' states in black hole candidates (BHCs) also identified by their timing and spectral properties. 
 
At a low luminosity state  
%level a BH transient   goes into outburst 
%it leaves a quiescent state and enters low hard state (LHS), a low luminosity
%state with 
the energy spectrum is dominated by a hard Comptonization component 
combined (convolved) with  a  weak thermal component.
The  spectrum of this low (luminosity) hard state (LHS)  
%to be believed 
is presumably a result of  Comptonization (upscattering)
of soft  photons, originated  in a relatively weak accretion disk,  off 
electrons of  the hot ambient plasma [see e.g. \cite{st80}].
Variability in LHS is high (fractional root-mean-square variability
is up to 40\%) and  presented by a flat-top broken power law (white-red noise) shape, accompanied by 
quasi-periodic oscillations (QPOs) in the range of 0.01-30 Hz, observed as narrow peaks in the power  density spectrum (PDS).
 In high soft state  (HSS) a photon spectrum is characterized by a prominent thermal 
component which is probably a signature of  a strong emission coming from a geometrically thin accretion disk.
A weak power-law component is also present at the level  
of not more than 20\% of the total source flux. In the HSS the flat-top variability 
ceases, QPOs disappear and PDS acquires a pure power-law shape.
The total  variability in HSS is usually about 5\% fractional rms.
The intermediate state (IS) is a transitional  state between LHS and HSS.
Note in addition to LHS, IS and HSS sometimes very soft state (VSS) is observed in which the BB  component is dominant and the power-law component is either very  weak or  absent at all. 
The  bolometric luminosity in VSS is a factor of 2-3 lower than that in HSS.

FB04 concluded  that probably all three states  A, B, C of GRS 1915+105 are instances
of something similar to the HSS/IS observed in other BHC systems, associated to the
high accretion rate value for this source, although during the hardest intervals 
LHS might be sometimes reached.
We come to the similar conclusions analyzing spectral and timing  data from GRS 1915+105 obtained by {\it RXTE} (see below).
 
Close correlations of nearly periodic variability 
[quasi-periodic oscillations (QPO)] observed during  low-hard and
intermediate states  with the photon index of the Comptonization
spectral component have been reported in multiple state transitions
 observed from accreting BHs [see \cite{vig}, \cite{ST06}, (2007),  (2009), hereafter V03,  ST06, ST07 ST09 respectively].
%Recently, the same type 
%of correlations have been firmly
%established for Cyg X-1 \citet{ST06},  for GRO J1655-40 \citet{sh07}.
The ubiquitous nature of  these correlations 
suggests that the underlying physical processes which lead 
to the observed variability properties are closely tied to 
the corona; furthermore, they vary in a well defined 
manner as the source makes a transition between spectral  states. 
The fact that the same correlations are seen in many sources, 
which vary widely in both luminosity (presumably with mass accretion rate)
 and state, suggests that 
the physical conditions controlling the index and the low frequency QPOs 
are characteristics of these sources.
% and that by 
%virtue of the low-high-frequency correlations \citep[see][]{BPK},
Moreover,  they  may be an universal property  of  all  accreting compact systems, 
including neutron sources too [see \cite{tf04}, hereafter TF04, and \cite{ts05}].

  When a BH is in LHS, radio emission is
also detected and a jet is either seen or inferred \citep{f01}.
Several models are successful in reproducing the energy
spectrum from the radio domain to the hard X-rays [see e.g.
\cite{MFF01}, \cite{vrc01}, \cite{cf02}, \cite{M03} and \cite{g05}]. The multiplicity of
models that can fit well the time average spectrum of galactic BHs indicates
that this alone is not enough to distinguish the most realistic one
among them. X-ray timing features can be  the key features to finally finding the common physical connection between the  corona, the accretion disk and the jet radio emission in BHs.

 There is a big debate in the literature on the origin of quasiperiodic oscillation (QPO) frequencies [see e.g. \cite{rm}] and its connection with the radio emission. \cite{mfk05}  reported on correlation between radio luminosity and X-ray timing features  in X-ray binaries containing  a number of low magnetic field  neutron stars and one black hole GX 339-4. They showed that in the low-hard state (LHS)   radio luminosity is correlated with the low frequency QPO  (LFQPO).  Note ST09 demonstrate that in  LHS of Galactic BHs    LFQPO  changes by order of magnitude, from 0.2 to 2 Hz whereas the photon index has almost the same value of 1.5.  Below we show that in GRS 1915+105 the photon index monotonically increases with 
 LFQPO and disk mass accretion rate, although  the radio luminosity does not correlate with LFQPO and X-ray luminosity in the whole range of spectral states, from low-hard to high-soft through intermediate states.
  Recently \cite{kyl08} suggested a model  which explains how  the QPO phenomenon is  related to an appearance of radio flares (jets).  Below (see \S 3) we present details of our observational study  of the QPO connection with the X-ray and radio flaring activity in GRS 1915+105.

%In the presented study we extend our study by considering 
%joint scaling of spectral index both in 
%QPO frequency and disk component normalization domains. 
In LHS and IS  which we consider in our study, only a
small part of the disk emission component is seen directly. The
energy spectrum is dominated by a Comptonization component presented by
a power law. To calculate the total normalization
of the ``seed'' disk blackbody component we model the spectrum
with a Generic Comptonization model [BMC XSPEC model, see details in \cite{bmc}] which consistently convolves 
a disk blackbody with a Green function of the Compton Corona to
produce the Comptonization component. We argue that the disk emission 
normalization calculated using this approach produces a more accurate
correlation with respect to the correlation with the direct disk component
which was obtained using the additive model, multicolor disk plus power law [see e.g.
\cite{mr}].
%Moreover, the disk normalization
%calculated in this way can be directly related to BH mass, emission efficiency 
%and system distance and geometry (see ST09).
%This allows to use the disk 
%normalization as a constrain of the BH mass and distance ratio for GRS 1915+105. 

%The main goal of this paper is to further study the correlations between photon index $\Gamma$ and QPO frequency in these sources  and their implications  to 
%BH mass determination in  galactic XRBs. 

%Thus, we use the  index-QPO and index-disk normalization 
%correlations to constrain the BH masses and distances for  
%Galactic BH sources H 1743-322,  GX 339-4,  XTE J1650-500, 4U 1543-47 
% and  XTE J1859+226  based on their expected dependence of 
%on the mass of the central BH, the source distance and geometry.   
%We use previously measured BH mass, system distance and inclination
%for XTE  J1550-564 as a reference information.
%An agreement of the result of our modified approach with masses 
%and distances for GRS 1915+105 given by
%the conventional dynamical,  spectroscopic  and scaling methods [see, \cite{o02}, \cite{st03} and ST07 respectively]  provide
%observational confirmation that basic assumptions behind the scaling of the index-normalization
%are correct.
This Paper is a continuation of the study of index-QPO and index-seed photon normalization 
correlations in BH sources started in  ST07 and ST09.  Particularly here we present a study of the   index-seed photon normalization (disk flux) correlation observed from GRS 1915+105 when it evolves from LHS to HSS.  The description of  \textit{RXTE} data-set used is given
in \S \ref{data}.  We have analyzed a broader sample of state transitions from GRS 1915+105
and  we found a diverse phenomenology for index evolution
through a transition.  In  \S \ref{obs_transition} we provide a detailed description of state transitions analyzed in this study.
In \S  4 we discuss and interpret the results of our observational study.
Specifically in  \S 4  we
consider the effect of the bulk motion Comptonization 
in the inner part of the accretion flow on the index evolution during a
state transition. Also we show that the index saturation 
effect is a direct consequence of the existence of this inner bulk motion
region and, therefore, can be considered as an observational signature 
of the converging flow (black hole). 
%In  \S 4.1 we also  discuss the transition of GRS 1915+105 to the thermal dominated state when the  modified 
%(by Comptonization) black body component prevails in the resulting spectrum.
%We infer the index of the Comptonization Green's function as a function of mass accretion rate  
%$\dot m$ (seed photon normalization). 
Furthermore in  \S 4 we  discuss  the TF04 model and  the Monte Carlo simulations  by  \cite{lt09} (in preparation) in which the observable index evolution with  $\dot m$  has been already  predicted.
%In the theoretical part (\S \ref{theory}) of the Paper we consider the physical 
%scenario which accounts for the observed phenomenology of 
%state transitions and also provide  explanation for
%the assumption behind the scaling of index-QPO and index-normalization 
%patterns. This is done in the framework of the transition layer (TL)
%paradigm, presented in \S \ref{tl}. In \S  
%\ref{inverse} we provide theoretical basis 
%for the inverse dependence of QPO frequency on BH mass 
%for a particular system state. In \S \ref{saturation} 
%There is a big debate in the literature on the origin of quasiperiodic oscillation (QPO) frequencies
%[see e.g. \cite{rm}].
%Recently \cite{kyl08} suggested a model in which the QPO phenomenon is closely related to an appearance of radio flares (jets).  In \S 3 we study in detail the QPO connection with the X-ray and radio flaring activity. 
% and find that during X-ray phase of the flare followed by that in radio the power spectrum is featureless, no QPO is observed (see \S 3).  
% In \S 5 we discuss and interpret  our observational  results.
 Conclusions
follow in \S \ref{summary}.

\section{Observations and Data Reduction \label{data}}

In the present Paper, we  have used publicly available data of
the \textit{RXTE} observatory obtained from January 1997 to April 2006. 
 In total, our study includes 107 observations made at
different  BH spectral states (LHS, IS, HSS) of the system.
Data sets were selected to represent a complete rise-middle-decay track 
of bright X-ray activity episodes behavior  along  bright radio
flaring events ($S_{15GHz}\ge$250 mJy). Therefore  we have chosen powerful 
($\ge$250 ASM counts/s) flaring episodes of GRS~1915+105
with a good coverage of simultaneous radio/X-ray observation.
% to demonstrate the index saturation effect in detail.
In the past some of these data of spectral transitions in GRS 1915+105  were analyzed by  \cite{trud99}, \cite{trud01}, \cite{muno99}, 
\cite{reig00},  ST07 and \cite{rodr08} for the  1997 -- 1998 and 2005 -- 2006 transitions respectively.

Standard tasks of the LHEASOFT/FTOOLS
5.3 software package were utilized for data processing  using
 methods recommended by RXTE Guest observer Facility according to 
``The RXTE Cook Book'' (http://heasarc.gsfc.nasa.gov/docs/xte/recipes/cook\_book.html).
For spectral analysis we used PCA {\it Standard 2} mode data, collected 
in 3 -- 20~keV energy range. The standard dead time correction procedure 
has been applied to the data. To construct broad-band spectra,
data from HEXTE detectors have also been used.
We  subtracted background corrected  in  off-source observations. 
%In order to account for the uncertainties in the HEXTE response and background determination
 Only HEXTE data  in  20 -- 150~keV energy range were 
used for the spectral analysis in order to exclude the channels with largest uncertainties. The HEXTE data have been re-normalized based 
on the PCA. The data are available through the GSFC public archive 
(http://heasarc.gsfc.nasa.gov). In  Tables 1-6 we list   groups
 of observations covering the complete dynamical range 
LHS-(IS)-HSS-(IS)-LHS 
of the source evolution 
 during flaring events. We present here period ranges MJD=50462 -- 51081 and 
MJD=53382 -- 53852, as  different types (samples) of bright X-ray activity, with 
%GRS~1915+105 
%outburst 
transitions between hard and soft states. 
Two selected data sets have  different patterns of radio/X-ray behavior    and of  light curve shapes. 
%of outburst properties.

We also use public GRS~1915+105 data from the  All-Sky Monitor (ASM) 
on-board \textit{RXTE} \citep{sw99}. The ASM light curves (2-12 keV energy range ) were 
retrieved from the public \textit{RXTE}/ASM archive at HEASARC (http://xte.mit.edu/ASM\_lc.html).

The monitoring {\it Ryle Radio Telescope} (15~GHz) data in the 1997 -- 2006 period were kindly provided by 
Dr. Guy Pooley. 
%(http://www.mrao.cam.ac.uk). 
The technical details of the radio telescope are described  by~\cite{pf97}.
 
%The integration is 5min per point.
%In normal conditions the noise level is about 2 mJy in this interval.
%Sometimes it is worse (weather, missing antennas).
%Averaging should reduce the noise in the expected way, though
%of course you might lose useful variability information.

%The data are described further in
%Pooley & Fender 1997 MNRAS 292 925
 
\subsection{Spectral analysis}
\subsubsection{BMC and iron line components of the model spectrum}
The broad-band source spectra were modeled in XSPEC with an additive model consisting of 
{\it two BMC}:
a BMC with high energy cut-off ({\it BMC1} component) 
and
{\it BMC2} component:
%an  
%absorbed 
%BMC1
% ($wabs*bmc$),  component {\it BMC1}, 
%and a BMC with high energy cut-off 
%($bmc*highecut$), component {\it BMC2}. 
%, component {\it BMC2}: 
$wabs*(bmc+bmc*highecut$). 
We also  use a multiplicative {\it wabs} model  to take into account of an absorption by neutral material. The {\it wabs} model parameter is an equivalent  hydrogen column $N_H$.
Systematic error of 1\% has been applied to the analyzed X-ray spectra.

The {\it BMC} model describes the outgoing spectrum as a convolution 
of the input ``seed'' blackbody-like spectrum, whose normalization  is
$N_{bmc}$ and color temperature is $kT$, 
 with the  Comptonization Green's function. Similarly to the ordinary {\it bbody} XSPEC model,
the normalization $N_{bmc}$ is a ratio of the source (disk) luminosity
to the square of the distance
\begin{equation}
N_{bmc}=\biggl(\frac{L}{10^{39}\mathrm{erg/s}}\biggr)\biggl(\frac{10\,\mathrm{kpc}}{d}\biggr)^2.
\label{bmc_norm}
\end{equation}
The resulting spectrum is characterized by the parameter $\log(A)$ related
to the Comptonized fraction $f$ as $f=A/(1+A)$ and a spectral index
$\alpha=\Gamma-1$.
There are 
several advantages of using the BMC model with respect to 
other common approaches used in studies of X-ray spectra of accreting
compact objects, i.e. sum of blackbody/multi-color-disk and power-law/thermal
 Comptonization. First, the BMC, by the nature of the model,
 is applicable to the general case where
there is an energy gain through not only thermal Comptonization but also via
dynamic (bulk) motion Comptonization \citep[see][for details]{ST06}.
Second, with respect to the phenomenological 
$powerlaw$ model, the BMC spectral shape 
has an appropriate low energy curvature, which is essential for
a correct representation of the lower energy part of spectrum.
Our experience with $powerlaw$ components shows that the model fit with 
this component is often inconsistent with the $N_H$ column values 
and produces an unphysical
component ``conspiracy'' with the $highecut$ part.
Specifically, when  a multiplicative component   $highecut$ is combined  with BMC,
%yields 
the cutoff energies $E_{cut}$ are in the expected 
range of 20$\sim$30 keV, while in a combination with
$powerlaw$, $E_{cut}$ often goes below 10 keV, resulting 
in unreasonably low values for photon index.
As a result, the implementation of the phenomenological 
$powerlaw$ model makes  much harder, or even impossible to 
correctly identify the spectral state of the source,
which is an  imminent task for our study. 
Third, and even a more important property of the BMC model,
it  calculates consistently the normalization of the original ``seed''
component, which is expected to be a correct mass accretion rate
indicator. Note  the Comptonized fraction is properly evaluated  by  the BMC
model. 

We consider a scenario related to our  spectral model (see Fig. \ref{geometry})
where the Compton cloud  along with converging flow  are located
in the innermost part of the source and the Keplerian disk is extended
from the Compton cloud (CC) to the optical companion (see e.g. TF04).
%(see  in F08  for the Comptonization model geometry).
An iron K$_{\alpha}$-line 
($laor$) component   \citep{laor} was included in our model spectrum. 
%The data require an iron K$_{\alpha}$-line ($laor$) component  for all IS observations. 
To summarize the  spectral model parameters  are
the equivalent hydrogen absorption column density {\bf $N_H$};
spectral indices $\alpha_1$,  $\alpha_2$ (photon index $\Gamma=\alpha+1$); 
color temperatures of the blackbody-like photon spectra  $kT_1$,  $kT_2$;
 $\log(A_1)$, $\log(A_2)$ related to the Comptonized fractions $f_1$, $f_2$ [$f=A/(1+A)$]; 
normalizations of the blackbody-like components {\bf $N_{bmc1}$}, {\bf $N_{bmc2}$}
for the {\it BMC1} and {\it BMC2} components of the resulting spectrum, respectively.
%high energy cut-off $E_{cut}$, and fold energy $E_{fold}$.

We find that color temperatures $kT_1$ and $kT_2$ are  about 1 keV 
for all available data  and thus  we fix values of $kT_1$ and $kT_2$ at  1 keV. 
%value of 1 keV in future calculations.
An equivalent hydrogen absorption column density was fixed at the level of 
$N_H=5\times 10^{22}$~cm$^{-2}$ \citep{trud01}.
{
When the parameter $\log(A)\gg1$ we fix  $\log(A)=2$   (see Tables ~1-6), because the Comptonized fraction 
$f=A/(1+A)\to$1 and variation of $A$  does not improve the fit quality any more.

During LHS  the BMC2 component is often very low or barely detectable 
%or undetectable at all 
[see Table 1 (MJD=50462-50561), Table 3 (MJD=51067-51081), Table 6 
(MJD=53829-53852) and Fig.~\ref{spec_evol_4Sm} and panel ``1S'']. This observational  fact is in agreement with scenario of BH spectral 
state transition (TF04). During LHS 
%$ E F_{E}$  diagram  (which units are 
%keV (erg cm$^{-2}$ s$^{-1}$ keV$^{-1}$)] peaks at about 100 keV  and 
the spectrum is characterized by a strong hard  power-law  component. In other words  the energy spectrum is dominated by a Comptonized component seen as a power-law hard emission in energy range from  $\sim 10$ to  $\sim 70$ keV while the disk emission remains weak (LHS, IS), because only a small fraction  of the disk  emission component $(1-f)$ is directly seen
(Fig.~\ref{spec_evol_4Sm},  panel ``1S'').

%uring LHS spectrum is peaked in the hard X-ray and 
%characterized by a strong power law (hard) component (Fig.~\ref{spec_evol_4S}, 
%panel ``1S''). 
 
%The energy
%spectrum is dominated by a Comptonized component seen as a hard power-law  
%emission.

Although during IS (Fig.~\ref{spec_evol_4Sm}, panel ``2S'') the contributions of  two BMC
components to the overall spectrum  are of the same order
% for the observable spectra 
sometimes  we can barely   identify one of these components  [see these cases in Table 2,
(MJD=50737, 50743), Table 3 (MJD=50908, 50909, 51003), Table 5 (MJD=53718) when 
N$_{bmc2}\ll$N$_{bmc1}$].
%(MJD=50737, 50743), Table 3 (MJD=50908, 50909, 51003), Table 5 (MJD=53718) 
%and panel S1 of Fig. \ref{spec_evol_4Sm}].
% the fitting procedure does not allow to distinguish BMC1 and BMC2 components.
On the other hand the model with two BMC components  are really needed  in the most cases of the intermediate state (IS) and high-soft state (HSS). In Figure \ref{two_BMC}  we demonstrate that the fit qualities are unacceptable when the only one BMC component is included in the spectrum.  Specifically,  for  IS-HSS  observation  91701-01-11-00 on 18 May 2005 the values of    $\chi^2_{red}$=12.3 for 75 d.o.f. (bottom left panel) for $wabs*bmc*highecut$ model. 
%  and $\chi^2_{red}$=12.1 for 77 d.o.f. (center bottom  panel)  for and $wabs*(bmc*highecut+laor)$ models  respectively.
However  $\chi^2$ is significantly improved when the second BMC component is included in the model.  
For  $wabs*(bmc+bmc*highecut)$ $\chi^2_{red}$=3.28 for  76 d.o.f.  (see the related count spectrum along with the model and the related residual in the central  bottom panel of Fig. \ref{two_BMC}). Ultimately we achieve a remarkable agreement with the data  using    $wabs*(bmc+bmc*highecut+laor)$ model  for  which $\chi^2_{red}$=1.04 for 73  d.o.f.  In Figure  \ref{two_BMC} (top and right bottom panels)   
we show the data along with  the best-fit spectra and their components for our two component BMC model (see Table 4 for the best-fit parameter values).
    
In HSS and very soft state (VSS) (MJD=53515-53640, Table 4) the soft luminosity is high and the
spectrum is dominated by a thermal component ($\Gamma_2>$3.7).  Note the hard power-law
 component is barely  seen in VSS (see Fig.~\ref{spec_evol_4Sm}, panels  ``4S'').
%For all this cases we used one BMC component to fit spectra.
%Herewith the normalizations define by the normalization of one BMC.
%The upper limits of normalization are 0.23 (LHS), 0.26 (IS) and 0.46 (HSS).
We find a broad emission line between 6 and 7 keV in the IS and HSS spectra.
Up to the date  the iron K$_{\alpha}$
emission line in IS and HSS of GRS~1915+105  was detected with ASCA, Chandra, XMM-Newton,
BeppoSAX by \cite{kot2000}, \cite{Mart02}, \cite{Mil05}. 
%But   there is some problem
%, using {\it RXTE} data, 
However the  determination of the iron line shape  with {\it RXTE} is a problem 
because of  the  low energy resolution.  
%  for RXTE 
%observation during very high X-ray emission level because the line is  broad and 
%hard to distinguish. 
As a first trial we added a Gaussian component 
%with a variable line 
 to fit the  spectrum varying the  width and normalization of the line.  Fits using the Gaussian   always produce   residuals  around 7 keV while fits with  XSPEC Laor model do not  have  such a  problem.
 %  a good quality fit.
 % of the line 
 Thus through  the Paper we incorporate  the Laor-line profile to fit the line component. The line
feature has a statistical significance of (3-10) $\sigma$ depending on the
spectral states. This line feature is variable with time average intensity of
2$\times$10$^{-11}$ erg/s/cm$^2$ and exhibits equivalent width  (EW) in the range of 50-600 eV across the data.
We found that adding the laor-line component significantly improves the fit quality of  
 IS and HSS spectra. Fitting an IS spectum
(e.g. for 90105-05-03-05 observation) without the line components leads to  
%theresulting spectrum with 
$\chi^2_{red}$=1.24.
%(for 72 d.o.f.). 
When  the   line component (Laor) is included
% the best-fit spectrum indicates 
the fit quality becomes much better,  $\chi^2_{red}$=1.01.
% (for 69 d.o.f.). 
The fit of HSS spectrum  (901050-08-02-00) without the iron-line  component is unacceptable,
%gives results with 
$\chi^2_{red}$=2.01,
% (for 72 d.o.f.) but   
%best-fit spectrum
%with 
and $\chi^2_{red}$=1.24 
%(for 69 d.o.f.) 
when the line component is included. 
%Figure 13  is the case  when the $\chi^2$ is larger than 1. The main
%scattering in this case revealed at high energies $>$60 keV. Also the small
%dipat 9-15 keV and low hamp near 20 keV are typical rest residual features.
%}
The  best-fit parameters of  the source spectrum and values of $\chi^2_{red}$ including d.o.f  
are presented in Tables 1-6.

%{\bf Lena, I want to make that this last statement is precisely true, is not it?  I hope  that you wrote in the previous version ``It is necessary to note that for  the spectral fitting we included 
%the systematic error 10\%'' is just typo.}.

%, where $\Gamma$ is a photon index. 
\subsubsection{Observational evidence of ``blackbody-like'' component peaked at $\sim 20$ keV in eight   IS spectra}  
The adopted spectral model shows a very  good performance for  99 cases among 107 spectra 
%all the data set 
used in our analysis. Namely, the value of reduced $\chi^2$-statistic
$\chi^2_{red}=\chi^2/N_{dof}$, where $N_{dof}$ is a number of degree of freedom
for a fit, is less or around 1.0 for most observations.  However for 8 observations of intermediate state 
%fraction (less than 3\%) 
the  fit  of the data with the model  {\it wabs*(bmc*highecut+bmc+  laor)} is not so good, 
%with high counting statistic
$\chi^2_{red}$ reaches 1.5 and even higher.   We found that in the residual of data vs model there is a characteristic bump  around   20 keV (see left bottom panel of Fig. \ref{sp_bbody}) which can be fitted by 
blackbody-like shape of color temperature about 4.5 keV (see right bottom  and top panels of Fig. \ref{sp_bbody} and Table 7 for values of the best-fit parameters).   This ``high-temperature BB'' component is  strong in each of the eight observations and its EW varies from 400 to 700 eV.  
We discuss a possible origin of this ``BB'' component   in \S 4. 
%However, it never exceeds a rejection 
%limit of 2.0.

\subsection{Timing analysis}
The \textit{RXTE}  light curves were analyzed using the {\it powspec} task. 
For the timing analysis in  2 -- 30 keV energy range,   we use the \textit{RXTE}/PCA data 
in the {\it binned} and {\it event} modes containing X-ray events below
13/15 keV and above 13/15 keV for 1997/2005 data sets respectively. 
Specifically,  depending on  {\it RXTE} epoch, the channel ranges (0-35) for binned and (36-255) for 
event modes relate   to energy bands 1.94-12.99 keV (binned) and 13.36-103.19 keV (event) for epoch 3 (1997-1998 data set), and  relate to 2.06-14.76 keV (binned) and 15.18-117.86 keV (event) for epoch 5 (2005-2006 data set).

The time resolutions for event 
%(E$_$16us$_$16B$_$36$_$1s) 
and binned  
%(B$_$8ms$_$16A$_$0$_$35$_$H$_$4P) 
modes are $1.52\times10^{-5}$ s and $8\times10^{-3}$ s, respectively. The observational  
exposition   time periods   vary from 1.5 to 10 ks. Thus for all of these observations  
 we can obtain  power spectra in the wide  frequency range (roughly 
from 0.001 Hz up to 100/10000 Hz for binned and event modes respectively). These  frequency ranges allow us to produce power spectra for all studied cases in  the  0.1-100 Hz frequency range. 
%We generated 
%power density spectra (PDS) in the 0.1 -- 100 Hz frequency range. 
%with sub-millisecond time resolution. 
We subtracted the contribution due to Poissonian 
statistics and Very Large Event Window for all power density spectra (PDS).

The data analysis of the PDSs was performed using  a simplified version of the  
diffusion model [see \cite{tsa07}, hereafter TSA07, and \cite{ts08}] in which the PDS continuum shape 
at frequencies below the driving frequency can be approximated by empirical model $P_X\sim (1.0+(x/x_{*})^2)^{-in}$ ($KING$ model in QPD/PLT). 
%a broken power-law ({\sc bknpl}).
%Recently \cite{tsa07} developed  and \cite{ts08} tested the physical  diffusion model for the power spectrum formation.   The underlying physical model of the physical  power spectrum is a white-red noise model is recently developed 
%and tested by X-ray observable power spectrum 
%by Titarchuk et al. 2007, 
%ApJ, 660, 556 and by Titarchuk & Shaposhnikov (2008), ApJ, 678, 1230.
Following TSA07, the break frequency found in the  PDS
is related to a diffusion time of  perturbation propagation 
while the QPO low frequency is an eigenfrequency of the volume 
(magnetoacoustic)  oscillation of the medium (in our case it is a 
Compton cloud). 
%Although the white-red noise (WRN)  model can be available as a working model which has been already used by TSA07 and TS08, it   can be approximated by the King shape as well. In another words the KING model takes into account the physical condition of origin PDS. 
Note, TSA07 demonstrated that the diffusion  model as a linear superposition of Lorentzians related to the eigenvalues of the diffusion problem can be also presented by the continuos  shape which is flat below break frequency and power law at frequencies above the break.  
Given these asymptotic forms  of PDS at low and high frequency limits they named their diffusion model as a
white-red noise (WRN) model.   For the quasi-uniform perturbation source distribution the slope of the PDS power-law part depends on the law of viscosity in the corona or in the disk (see details in TSA07). The parameters of  this WRN diffusion model are the break  (diffusion) frequency and the index of the power-law distribution of the viscosity over the radius. 
%To model the PDS we used the QPO/PLT plotting package.
%The PDS continuum shape in the LHS and IS usually has
%a band-limited noise shape, which is well represented by an 
%empirical model $P_X\sim (1.0+(x/x_{*})^2)^{-in}$
%($KING$ model in QPD/PLT).
%The parameter $x_*$ is related to the break frequency and
%$2\,in$ is the slope of the PDS continuum after the break. 
% We use recently a developed perturbation diffusion model
%\citep{tsa07} to fit PDS continuum.
To fit the QPO features, we use Lorentzian shape.
We quote the Lorentzian  centroid as a QPO frequency.

\section{Observational results  \label{obs_transition}}

\subsection{Evolution  of spectral  properties during
 state transitions \label{transitions}}

Observations of Galactic BH X-ray binaries reveal 
diverse spectral  and dynamic phenomenologies. The evolution of a BH binary
%in outburst 
is usually  described in terms of spectral states. 
There are
several flavors of BH state definitions in literature, which
slightly differ in BH state definitions and terminology
\citep[see, for example][]{rm,bell00,bell05,kw08}. To distinguish
different states, the properties
observed in the energy spectrum and 
Fourier power density spectrum (PDS) are utilized.
As we have already emphasized in the introduction section  we use, in our study,  the general  state classification for 
four major BH states:
%{\it quiescent}, 
{\it low-hard} (LHS), 
{\it intermediate} (IS), {\it high-soft} (HSS) and {\it very soft} 
state (VSS). 
%When a BH transient  goes into  outburst 
%it leaves a quiescent state and enters the LHS, a low luminosity
%state with energy spectrum dominated by a power law component 
%and characterized by a very weak thermal component. Efficient  Comptonization
%of photons originated in a weak accretion disk and scattered  off 
%electrons in the hot ambient plasma is believed to be responsible
%for the formation of the energy spectrum in LHS [see e.g. \cite{st80}].
%The variability in the LHS state is high (fractional root-mean-square variability
%is up to 40\% ) and is represented by a white/red noise broken power law, accompanied by 
%quasi-periodic oscillations (QPOs), seen as narrow peaks in PDS (TSA07).
 %For strong outbursts, after several days or weeks, BHs enter into 
%a HSS, when the spectrum is presented by a thermal 
%component, attributed to a strong emission from a thin accretion disk.
%A weak power-law component is also present at the level of 
 %not more than 20\% of total source flux. The IS is a transitional 
%stage between LHS and HSS. In this state   the energy spectrum 
%becomes softer than that in LHS and a  fraction of Comptonized photons   drops.
%The aperiodic variability fraction also decreases, which is accompanied 
%by an increase of all characteristic frequencies in the PDS.

The general picture of LHS-IS-HSS transition is illustrated
in Figure \ref{spec_evol_4S} where
%In Figure \ref{spec_evol_4S} 
we bring together spectra of LHS, IS, HSS and VSS  to demonstrate 
the source spectral evolution from low-hard to soft states.    
%We discuss more details of the spectral transitions and their properties in \S 3. 
We should emphasize different shapes of the spectra for  the different spectral states.  In the LHS spectrum the Comptonization component is dominant   and  the blackbody (BB) component is barely seen  in  3-150 keV energy range. The IS and HSS spectra are characterized by a strong soft BB component  and a power law extended up to 150 keV.
In  VSS  the soft BB component is dominant and the power-law component is relatively weak with respect of this in IS and HSS.
%\ref{spec_evol_97_rise} 
%where we present the energy spectral diagrams $E\times F(E)$ 
%(in units of (keV)$^2$/cm$^2$/s/keV).
 
%of three representative \textit{RXTE} observations of GRS  1915+105
%from LHS, IS and HSS during 1997 rise outburst. 

{ In the RXTE data of GRS 1915+105 observations there are long periods when the photon index 
$\Gamma_1$ and normalization $N_{bmc1}$ of the hard BMC monotonically increase (or decrease) with time.
We call these periods  as long 
%outburst 
transition periods. The days,  when the source X-tay flux  starts to increase while it is still  
%to leave
%the quiescent state and 
%to enter 
in the LHS,   can be considered  as  a  beginning  of the rise
% outburst 
transition.   
%BEGIN of seif add
%In this Paper we adopted the follow definition of outburst transition limits.
In these times the energy spectrum is characterized by  low index
values  $\Gamma_1\sim$1.7 and the thermal component  is at low level
%(with ($\Gamma_2\sim$3.0)) 
or  undetectable at all. 
In Figure \ref{outburst_97_rise}
% -- \ref{outburst_97-98_decay} 
from top to bottom
we show an evolution of flux density $S_{15GHz}$ at 15 GHz (Ryle Telescope), 
\textit{RXTE}/ASM count rate, 
BMC normalization and 
photon index $\Gamma$ 
during the 1997 
% outburst 
rise transition of GRS~1915+105 (MJD 50500 -- 50700).
Red/black points ({\it for  two low  panels}) correspond 
to hard/soft components with $\Gamma_1$ and $\Gamma_2$, respectively. 
In the bottom we plot the photon index $\Gamma$ versus the BMC normalization ({\it left}) 
and Comptonized fraction ({\it right}) for this  transition. Here red 
triangles/black circles correspond to hard/soft components with 
$\Gamma_1$  and $\Gamma_2$, correspondingly.
One can see that in the beginning of this transition the resulting spectrum consists of one 
Comptonization component whose photon index $\Gamma_1$ steadily increases {\it from 1.7} towards 
the softer states and finally
saturates at the value of 3. The Comptonization fraction $f=A/(1+A)$ of the hard component, related
to index $\Gamma_1$, shows a sign of decreasing towards the softer state.
When the \textit{RXTE}/ASM count rate exceeds 50 counts/s  the soft Comptonized component
appears with a  weight  which is  comparable to that of the hard component. The photon index of the soft component $\Gamma_2$ saturates to the 
level of 4.2 when the BMC normalization (disk flux) increases. The Comptonization fraction $f$ of the soft component is about 0.5 and higher.

As seen from Fig. \ref{outburst_97_rise} the start of this  rise transition  coincides  with active phase of X-rays, 
and   of radio emissions  which  exceed  10 ASM counts/s   and 50 mJy  levels respectively in the 1997 rise transition. Around  MJD 50580 day the source reaches the HSS  (when $\Gamma_1\sim 3$). 
% the power-law index  of the hard component  (BMC1)  $\Gamma_1$ saturates at 3. 
Then   a long HSS period from MJD 50580  to 50700,  when $\Gamma_1$ stays almost the same,  
is followed by the state transition to IS during which  $\Gamma_1$ decreases to 2.5  
(see Fig.   \ref{outburst_97_middle}-\ref{outburst_97-98_decay}).

We see a similar behavior of X-ray,  radio fluxes and photon indices during the 2005  bright X-ray  episode. The only difference of that with the 1997 episode was that  the 2005 rise 
%outburst 
started at the intermediate state and went very  quickly to HSS  (see Fig.   \ref{outburst_05_IS}-\ref{outburst_05-06_decay}). After  MJD 53800 the source came back to IS-LHS when $\Gamma_1\lax 2$
(see Fig.  \ref{outburst_05-06_decay} and Table 6). Note typical X-ray  and radio fluxes during  IS  are about   40-60 ASM counts/s and    $\le$50 mJy respectively.

\subsection{
%Correlated and non-correlated 
Observational (correlated and non-correlated) characteristics of X-ray and radio emissions} 
In fact, we do not find any correlation of X-ray and radio fluxes when the source changes its spectral states. Also we do not find a correlation of radio activity with the X-ray photon index (see Fig. \ref{outburst_index_radio_X-ray}). However we find a strong correlation  of the iron line EW with radio flux density  $S_{15G{\rm Hz}}$ at 15 GHz (see Fig. \ref{outburst_EW_radio}). 
In  Figure \ref{outburst_EW_radio} we also include points which have been recently reported  by \cite{nl09}  who have analyzed archival HETGS (High Energy Transmission Grating Spectrometer) observations of GRS 1915+105 from the Chandra X-ray Observatory. 

}

{
%\it Radio flux sometimes can stay at the low level ($\le$50 mJy) stil 100 days (1997 set).
}

%The dominance of  the soft component  in the resulting spectrum  (so called VSS) is very uncommon in the GRS 1915+105   outbursts.  
%Although 
The prominent HSS events were observed  during 2005 -- 2006 observations
around MJD 53490 and MJD 53690  (see Figs. \ref{outburst_05_IS}-\ref{outburst_05-06_decay}). 
The 2005 -- 2006 observations confirm the index evolution vs BMC normalization (disk flux) found in the 1997 -- 1998.  The indices of  the hard and  soft components indeed increase and then saturate  
at values  of $\lax 3$ and 4.2 respectively (see Fig. \ref{outburst_05-06_decay}). 
%{it 
%As seen from this Figure  
Index $\Gamma_1$   started  a saturation  at lower values of  BMC1 normalization  (presumably  proportional to disk mass accretion rates) than that were   in 1997. 
%the saturation for the hard BMC i happens .  
%These points belong to IS plateau and reasonably to connect the physical
%interpretation with special physical (or geometrical) conditions of for the
%hard BMC to saturates  during more lower than at HSS and constant X-ray luminosity. 
%Note It is importat that radio flux  can significantly increase/decrase during constant X-ray
%luminosity of IS with saturation of spectral index effect.
%}

 It is also worth noting a so called  ``pivoting'' effect, 
i.e. when  inequality   $N_{bmc1} >N_{bmc2}$ 
switches to   $N_{bmc2} >N_{bmc1}$ 
%$BMC2-\rm{norm} >BMC1-\rm{norm}$ 
or vice versa, which is 
seen in the 1997-1998 and 2005 - 2006  observations. One  can see  this composite pivoting  picture combining  Figs. \ref{outburst_97_rise}-\ref{outburst_05-06_decay}.  In fact, these pivoting points
correspond to  the spectral transitions between adjacent  states  LHS-IS to HSS and vice versa.
% HSS to IS-LHS.

In Figure \ref{outburst_index_norm} we collect all data points for the index-normalization correlation 
for rise and decay 
%outburst  
stages. We  do not find much differences in  the correlation patterns 
related to the  rise and decay transitions
% outbursts 
(compare left and right panels) in contrast  that  ST09 found  in other BHs. 
%Given that we put together all correlation points of the hard component index $\Gamma_1$ vs BMC normalization in Fig. \ref{scaling_index_norm} (see red points there).

We also find that the photon index of X-ray spectrum is tightly correlated with 
the quasi-periodic oscillations  (QPO) frequency 
(see Fig. \ref{outburst_index_qpo}) which can be considered as a strong argument that QPOs and X-ray Comptonization spectrum  emerge from the same geometrical configuration  (Compton cloud).
However the flux density $S_{15GHz}$ and QPO frequency are not correlated with each other  
when the source changes its spectral states. 
In Figure \ref{radio_QPO_independence}  we show
 an evolution of the flux density $S_{15GHz}$ at 15 GHz (Ryle Telescope), 
\textit{RXTE}/ASM count rate and  $\nu_{QPO}$ 
during 1997 (left column) and 2005 (right column) 
%outburst 
rise transitions. 
%Here $f_{QPO}$ denotes the centroid frequency of the fundamental QPO. 
The left column panel demonstrates the presence of QPO during a low radio activity  ($<$30
mJy).  The right column panel shows an example of the presence of QPO when the radio flux is high 
($\sim$100 -- 200 mJy). Given that the quasi-periodic oscillations  (QPOs) of X-ray flux are present 
independently of the radio flux level, we can conclude that  the radio appearances and QPO phenomenon  are not closely related and probably  the  radio and X-ray (oscillating) emission  areas   have different origins  in the source.
% [cf. \cite{kyl08}].

\subsection{Evolutions of energy and power spectra during a minor X-ray/radio flares \label{minor_flare}}
In Figure \ref{radio_appearances} we show the details of a typical  evolution 
of X-ray timing and spectral characteristics for minor X-ray/radio flares.
% outbursts.
%of the flux density $S_{15GHz}$ at 15 GHz (Ryle Telescope) and 
%of the \textit{RXTE}/ASM count rate during the rise transition stage of the  2005 outburst.
%{
%It is note that as consistent with definition of (limits of) outburst transition, the END of
%outburst corresponds to decreasing of $Gamma_1$ to $\le$1.7, when object return to
%quiescent state. Thus, the apparent final of 2005 outburst at MJD=53710 is not a real final.
%In spite of X-ray emission droped to 50 ASM cnts/s and radio flux level decreased to
%$\sim$80-90 mJy (Fig. 7), the index $\Gamma_1$ was around 2.7 yet. So, the 2005
%outburst was in progress yet. In point of fact (definition of transition END), the 2005
%outburst really finished at the MJD=53830 only, when the index $\Gamma_1$ declined to 1.7.
%}
%  from GRS~1915+105.
In the top  panels of Figure \ref{radio_appearances}  we show  the flux density $S_{15GHz}$ at 15 GHz (Ryle Telescope) and   the \textit{RXTE}/ASM count rate during the 2005 rise transition stage
% of the   outburst 
(see also Fig. \ref{outburst_05_IS}).
Red points A, B and C on the panel of the \textit{RXTE}/ASM count rate vs time correspond to    
moments at MJD=53416, 53422 and 53442 (before, during and after the minor X-ray/radio flare)
respectively.
%{
Points A and C were chosen as  the nearest possible points   to point B (taking into account 
the time-table of archive data). Point B  corresponds
to the maximum  of  radio
 flux of 300 mJy and EW of 600 eV.   % at short HSS. While points A and C belong to IS. 
Note  that QPO centroid frequency before the  flare (at point A) is  at 1.8 Hz and  
shifts to 0.9 Hz (point C) after the flare. 
%Thus the object is some greater 
%after flash. Further, during IS plateau the 
%QPO centroid low shift (increase) to 10 Hz with decreasing of amplitude and than 
%distructe at HSS flash with very high soft X-ray. Thus the object is some clamped 
%before X-ray outburst.
%}

 PDSs  (left bottom column) are plotted versus the energy spectrum 
(right bottom column) for three points A (top), 
B (middle) and C (bottom) of the X-ray light curve. There are QPOs at A and C points 
(A1, C1 panels) but there is none at   B point   (B1 panel), at the X-ray flare peak. 
%Components of  spectrum [3-20 keV] demonstrate rising of soft component 
%and Fe K-line complex contribution during radio/X-ray outburst. 
For the photon spectra (right bottom column) red points stand for observational data, while the  model  is shown by  components 
with blue line for  {\it BMC1},  black line for 
{\it BMC2}  and dashed purple line for the 
{\it laor} line component.
Note that the spectral characteristics  undergo noticeable  changes during X-ray/radio  
flare.
%outburst.  
Specifically  at the flare peak   (point B) the total flux  increases at least by  a factor  of 1.5  with respect to  that before the flare.   
%(small X-ray outburst), when object enter to short HSS at point B
Although photon index of {\it BMC1} component  $\Gamma_1$ changes from 2.9 
(A and C points) to 3.0 (B point) respectively. 
%While the X-ray flux  at the peak (point B) 
%is at least a factor  of 1.5 higher than that  before and after outburst (A and C points).
%his point the equivalent width EW increases up to 600 eV.

We also studied the energy dependence of the PDS shape and integrated power variability as a function of
the photon energy.  In Fig. \ref{radio_appearances} ({\it left bottom column pannel}) we show 
two power spectra for  two energy bands 
2-15 keV (red) and 15-30 keV (blue). 
One can see that PDSs  weakly depend   on the energy band.
% in the high energy band   (15-30 keV)
%the variability is much higher  than that in  the low energy band   (2-15 keV). Also we should note that
% the low energy PDS   can be fit by the white/red noise (WRN)  shape,  with break frequency at about 2 Hz. 
% One can see that  QPO Lorentzian is centered  at 2 Hz too. While the high energy PDS is fit by  the white noise power-law of index 0 along with the QPO component  at 2 Hz.  
 In particular, a  value of  the low frequency QPO $\nu_{QPO}$
 is the same for    the low energy and high energy  PDSs.
 %   is present at 2 Hz and that of  the high energy PDS is not seen up to frequency 20 Hz 
%According to \cite{tbw01} 
%We can suggest that the sizes of the photon emission areas $L_{CC}$ related to these two energy bands are also the same. In fact,  \cite{tbw01} argue that  $\nu_{QPO}$ is proportional to the ratio of magneto-acoustic (plasma) velocity  $V_{MA}$ and Compton cloud size   $L_{CC}$ and thus one can conclude that
 %Thus  if $V_{MA}$ is determined by a spectral state only 
 %Then 
 %the emission areas  are the same  because  $\nu_{QPO}$ are the same  for these two energy bands.

%Both panels demonstrate the independence of the QPO
%appearance from radio events of GRS~1915+105. 

\section{Interpretation and discussion of observational results \label{theory}}
Before to proceed with the interpretation of the observations let us to  briefly summarize them as follows:
i. The spectral data of GRS 1915+105 are well fit by two  (soft and hard)  BMC components for most of analyzed IS and HSS spectra (see Fig. \ref{two_BMC}) while LHS spectra essentially   require only one BMC component, the soft BMC component is very weak (see Tables 1, 3-4, 6 and  panel S1 in Fig. \ref{spec_evol_4Sm}).
ii. In addition to two BMC components  8  IS-HSS spectra require  an extra component which can be fitted by `` high temperature BB-like"  profile (see Fig. \ref{sp_bbody} and Table 7).
iii.  The Green's function indices of each of these components rise and saturate with an  increase of the BMC normalization (disk flux). The photon index saturation levels of the soft and hard components are about 4.2 and 3 respectively (see Fig. \ref{outburst_index_norm}).  iv. We also  find  a tight positive correlation of QPO frequencies with the index  (see Fig. \ref{outburst_index_qpo}) and consequently that with the disk flux. vi. The  iron line EW correlates with the radio flux 
(see Fig. \ref{outburst_EW_radio}). vii. QPO appearances  and their frequency values are not correlated with radio flux when the source undergoes the spectral changes from IS to HSS 
(see Fig. \ref{radio_QPO_independence}).
vii. We also do not find any correlation between X-ray and radio fluxes  and X-ray power-law index
(see Fig. \ref{outburst_index_radio_X-ray}). 
viii. Although we find  changes of power and energy spectra  during a minor X-ray/radio flares when 
QPO features disappear   in PDS and energy spectrum becomes softer than that was   before and after the flare (see Fig. \ref{radio_appearances}).
%Note that we find   only a few LHS spectra  $\Gamma_1\lax 1.7$ in GRS 1915+105 and therefore we can or refute of QPO-radio correlation early found in LHS of GX 339+4  

\subsection{Index-QPO  and index-$\dot m$ correlations. Index saturation}
We confirm the  index-QPO correlation in GRS 1915+105 previously found by V03 and ST07. 
This correlation was indeed predicted by \cite{tlm98}, hereafter TLM98, who argued that the transition layer  [Compton cloud (CC)] formed  between the Keplerian disk and the central object (NS or BH), contracts and becomes cooler when the disk mass accretion rate $\dot m$ increases.  
The observational effect of the CC contraction were later demonstrated by ST06, TSA07, TS08 and 
\cite{mtf09} in Cyg X-1 and XTE J1650-500 respectively.
% When   $\dot m$ increases  
 As a result of the transition layer (TL) contraction the QPO low frequency  $\nu_L$,  which is presumably the TL's  normal mode oscillation frequency,  increases with $\dot m$ 
given that $\nu_L$   is inversely proportional to the TL size. On the other hand the index monotonically increases when  the TL (CC)  cools down. TF04 provided the details of the index-QPO correlation model where they pointed out  that this correlation is a natural consequence of the spectral state transition.
 % {QPOthe transition layer (Compton cloud) becomes more compact and cooler with respect to this in LHS. 
%TF04 argue that the  index-QPO correlation is naturally expected 
 %and the index monotonically increases with the TL plasma  temperature  

In this Paper we have firmly established the index correlation with $\nu_L$ along with the index  saturation vs the BMC normalization $N_{bmc}$  (Eq. \ref{bmc_norm}) for the soft and hard Comptonized components of the X-ray spectra of GRS 1915+105 (see Fig. \ref{outburst_index_norm}). Below we   show that 
$N_{bmc}$ is actually proportional to mass accretion rate in the disk.
Namely  the disk flux $L$ (as a source of soft photons for Comptonization, see e.g.  Fig. \ref{geometry} for the geometry of soft photon illumination of Comptonized region)  can be represented as 
\begin{equation}
L=\frac{GM_{bh} \dot M}{R_*}=\eta(r_*) \dot m_d  L_{\rm Ed}.  
%\sim \frac{GM_{bh}}{R_S} \dot{M} \eta \sim   \dot{M}c^2 \eta = M_{BH} \dot{m} \eta.
\label{lumin}
\end{equation}
Here $R_{*}=r_{*}R_{\rm S}$ is an effective radius where the main energy release takes place in the disk,  $R_{\rm S}=2GM/c^2$ is the Schwarzschild radius, $\eta=1/(2r_*)$, $\dot m_d=\dot M_d/\dot M_{crit}$ is dimensionless mass accretion rate in units of the critical mass accretion rate
$\dot M_{crit}=L_{\rm Ed}/c^2$ and $L_{\rm Ed}$ is the Eddington luminosity.

On the other hand 
\begin{equation}
L_{\rm Ed}=\frac{4\pi GMm_pc}{\sigma_{\rm T}}
\label{ed_lumin}
\end{equation}
i.e. $L_{Ed}\propto M$ and thus using Eqs. (\ref{lumin}-\ref{ed_lumin})
we obtain that 
\begin{equation}
L\propto\eta(r_*) \dot m_d m.
\label{lumin_m}
\end{equation}

In HSS when the inner disk radius $R_{*}$ reaches its  lowest value  $R_{*}\gax3R_{\rm S}$, 
the efficiency of the gravitational energy release $\eta(r_*)$ reaches its highest value  and thus 
the disk flux increases only when  the disk mass accretion rate increases (see Eq. \ref{lumin_m}).
Given that BMC normalization $N_{bmc}$ is proportional to $\dot m_d$  in HSS
 the observational effect of the index saturation with   $N_{bmc}$ is translated to
the index  saturation with  $\dot m_d$.

First we interpret the index saturation related to the hard Comptonization component ({\it BMC1}).
We suggest that this BMC1 component of the emergent spectrum is presumably originated in the converging flow onto a compact object, in our case to the BH 
(see Fig. \ref{geometry}). 
In fact,  in  HSS the plasma temperature of the accretion flow is comparable with the color temperature of the disk photons (see TF04). 
Thus, in order to explain
the high energy tail observed in HSS of BH sources,  one should assume either an unknown source of high energy non-thermal electrons 
[see e.g. \cite{cop99}] or consider effects of energy transfer from the
 converging flow electrons to the photons 
 %\textbf
 {emitted} from the innermost part of the accretion flow.
 
Optical depth of the converging flow $\tau$  is 
proportional to  $\dot m_d$ if one assumes that  disk accretion flow continuously   goes to the converging flow and there are no other components in the accretion flow [see e.g. a model of  two component accretion flow by \cite{CT95}, hereafter CT95].
% [see \cite{bmc} and \cite{tz98}, hereafter TZ08].
This effect of the index saturation vs optical depth of the bulk flow (BM) $\tau$  was first predicted by \cite{tz98}
%hereafter TZ08    
and then it was subsequently reproduced in Monte-Carlo simulations by \cite{lt99}, hereafter LT99.  

It is worth noting that the index saturation effect is an intrinsic property of the bulk motion  
onto a BH   given that the spectral index $\alpha=\Gamma-1$ is a reciprocal of the Comptonization parameter 
$Y$ [see  this proof in  ST09 and \cite{BTK07}] which  saturates when the BM optical depth, or $\dot M$, increases.  
In fact, the Y-parameter is a  product of  the average photon energy exchange per scattering $\eta$ and the mean number of photon scattering $N_{sc}$, i.e. $Y=\eta  N_{sc}$.
For the thermal Comptonization case, $Y\sim (4kT/m_ec^2)\tau^2$  given that in this case  
$\eta=4kT/m_ec^2$ and $N_{sc}\sim \tau^2$ for $\tau\gg 1$   \citep[see e.g.][]{rl79} 
   and, thus, the thermal Comptonization spectral index is
\begin{equation}
\alpha\sim [(4kT/m_ec^2)\tau^2]^{-1}.
\label{alpha_plmm}
\end{equation}
In the case of converging flow,  
%for $N_{sc}\gg1$   
the preferable direction for upscattered photons is    the direction of bulk motion onto the BH,  i.e along the radius.
Note that the fractional photon energy change is 
$$
\Delta E/E=(1-\mu_1 V_{R}/c)/(1-\mu_2 V_{R}/c).
$$ 
where $\mu_1$ and $\mu_2$  are the cosines  of the angles between the direction of the electron velocity ${\bf n}={\bf V}_R/V_R$  and direction 
of incoming and outcoming (scattered) photons respectively. 

%Given that  $\Delta E/E$ has a maximum at $\mu_2=1$ for given $\mu_1$ and $V_R$
  The number of scatterings of the up-Comptonized photons $N_{sc}$  can be estimated as a ratio of the radial characteristic size of the 
converging flow  $L$ and the free path $l$ in the direction of motion, namely $N_{sc}\propto L/l=\tau$
given that  $\Delta E/E$ has a maximum at $\mu_2=1$ for given $\mu_1$ and $V_R$.
On the other hand  the efficiency per scattering  for bulk motion flow $\eta\propto 1/\tau$  when 
$\tau\gg 1$ [\cite{lt07}, hereafter LT07]
 %\textbf
 {hence for bulk motion Comptonization, the Y-parameter does not 
depend on $\tau$} for high values of $\tau$ or dimensionless mass accretion rate $\dot m$.
Thus one can conclude that  {\it the Comptonization parameter $Y=\eta N_{sc}$  and  
 hence the energy index $\alpha=Y^{-1}$}   {\it saturate to a constant value when optical depth (or mass accretion rate) of the BM flow increases}.
 
 However the index saturation  value is determined 
 by the plasma temperature  during a  transition [see LT99].  
  The plasma temperature strongly depends on the  mass accretion rate in the 
bulk motion region $\dot M_{bm}$ 
and  its illumination by the disk photons $L$  (see TLM98 and TF04). For higher $\dot M_{bm}$ and  $L$  the 
plasma  temperature is lower.  The level of the index  saturation decreases when  the plasma 
temperature in the bulk motion increases (TF04). 
Thus the index  saturation levels can be  different from source
 to  source depending on the strength of the disk. 
 Looking at Figure \ref{outburst_index_norm}  one can also notice that the index $\Gamma_1$ starts its saturation at different values of BMC normalization ($\propto\dot m_d$) for different  types of active episodes.
 %outbursts.  
 In fact, the index should saturate with mass accretion rate in the converging flow $\dot m_{bm}$ which is a sum of the disk mass accretion rate $\dot m_{d}$   and that in  sub-Keplerian flow (CT95).  
Hence one can argue that this lower value of $\dot m_d$  at which  the index saturates can be  a sign  of the presence of extra (sub-Keplerian) component  in the accretion flow onto BH in GRS 1915+105.

 \cite{lt09}, hereafter LT09,
study the index-$\dot m$ correlation and also a modification of the disk blackbody spectrum due to Comptonization  in the optically thick  CC, which is formed due to accumulation of  accretion matter in the TL. 
% between the  inner  sonic radius and the radius of an adjustment of TL to Keplerian disk.   
They indeed explain the saturations of the indices  of the soft and hard components of the resulting spectrum.  Specifically  LT09 show that gravitational energy of the accretion flow is released in the optically  thick and relatively cold TL when mass accretion rate $\dot m_d$ is higher than 1. 
%The exact value of  $\dot m_d$ depends on the TL layer model parameters.  
The level of the index saturation depends on the radial velocity in the transition layer (TL).  
LT09 also show that the observable saturation index  of the soft BMC component  $\Gamma_2\sim 4.2$ can be reproduced in their Monte Carlo  simulations   for  values of the TL radial velocity $\gax 0.05$ c.   

%It is remarkable that independently from the model parameters, 
%the photon  index of Comptonization Green's (scattering spread) function of this Compton cloud (CC) saturates to  constant value of 4.2.   In our observational study of X-ray spectra from GRS 1915+105 we find a confirmation of the LT09 prediction regarding the index saturation of the BB modified spectral component  ({\it BMC2}). From Figure \ref{outburst_index_norm}
 %one can clearly see that the  {\it BMC2} index $\Gamma_2$ saturates to a constant of 4.2 when 
 %the disk mass accretion rate (disk flux) increases.   
\subsection{Physical origin of  ``high temperature BB-like'' component''?} 
In 8 of  IS-HSS spectra   we find  an observational evidence of the bump around  20 keV which  can be fitted by ``$\sim 4.5$ keV BB-like'' profile (see Fig. \ref{sp_bbody} and Table 7). One  can argue that this observable bump at 20 keV is a signature of the Compton reflection bump 
[see e.g. \cite{bst74}, ST80, CT95, and  \cite{mz}].   But this interpretation encounters difficulties
%But there is a big problem with the Compton bump interpretation  
given that the hard power-law tails of these spectra  are too steep to form the Compton bump. Indeed, ST80 and later LT07 demonstrated that the Compton bump as a result of  a photon accumulation due to downscattering  of hard photons in the cold medium (for example disk)  cannot be produced  if the photon index of the incident hard photon spectrum  $\Gamma>2$. In fact, as one can see from Table 7,  that in all spectra where we detect this 
$\sim 20$ keV feature the index of the hard BMC component $\Gamma_1>2$  
(the indices vary  between    2.5 and 3).

In principle, this bump may also be a result of photoelectric absorption of the  photons below 10 keV in the cold medium  (disk) even if the incident spectrum is very steep. The photoelectric absorption cross-section $\sigma_{ph}\sim (7.8~\rm keV/E)^3\sigma_{\rm T}$, where E is photon energy and    
$\sigma_{\rm T}$ is Thomson cross-section (e.g. CT95).   However, \cite{lat04} and \cite{rm08} show that  the ionization of such a disk by the intensive  X-ray radiation during IS-HSS invalidates the basic assumptions of the presence of the cold material  in the innermost part of the source. Note the hard tail of X-ray spectrum in IS-HSS is formed in the converging flow (CF), i.e. in the innermost part of the accretion flow,  because we see the CF signature (the index saturation) when the source goes to IS-HSS.
%might invalidate some of basic assumptions

On the other hand \cite{t02} argued that the specific  spectral and timing features of X-ray radiation 
 could  be seen in BH sources  only. Particularly, he stated that
% that photon-electron interaction along with the general relativistic effects lead to the formation of a specific spectrum. 
the photon-photon interaction  of the effectively upscattered photons results in the powerful pair production near a BH  horizon.  Indeed, a large fraction of the  upscatterred photons going inward are deflected by the relativistic free-fall electron in the outward direction [the aberration effect of light, 
see e.g. \cite{rl79} and Appendix A]. 
%As a result 
These diverted upscattered   photons of energy $E_{up}$ interact with incoming photons
of energy $E_{in}$ flux and ultimately this interaction leads to the pair creation if the condition $E_{up}E_{in}\gax (m_ec^2)^2$ is satisfied.    
Note  that  free-fall bulk  motion with  Lorentz factor $\gamma\gg1$ 
should be very close to horizon, i.e.
\begin{equation}
\frac{\Delta R}{R_{\rm S}}\approx \frac{1}{\gamma^2}
\label{distance_to_horizon}
\end{equation}
where $\Delta R=R-R_{\rm S}$ is radial distance to horizon. Thus 
the created  positrons extensively  interact with accreting electrons there and therefore the annihilation line photons are created and distributed over  the relatively narrow shell near  BH horizon. Specifically 
\begin{equation}
\Delta R \lax3\times10^{4}\left(\frac{10}{\gamma}\right)^2\frac{m}{10}~{\rm cm}.  
\label{distance_to_horizon_m}
\end{equation}
The proper energy (in the comoving flow frame) of annihilation line photons $E_{511}$ should be seen by the Earth observer  (in the zero frame) at the redshifted energy 
%(see Eq. \ref{distance_to_horizon})
\begin{equation}
E_0=(1-R_{\rm S}/R)^{1/2}E_{511}\approx\frac{E_{511}}{\gamma}
\label{redshift}
\end{equation}
where $(1-R_{\rm S}/R)^{1/2}\approx(\Delta R/R_{\rm S})^{1/2}=1/\gamma$ (see according   
Eq. \ref{distance_to_horizon}).
%Note  is, by definition,

In other words  the line energy displacement due to gravity as viewed by a far away observer in free space is $z+1=1/(1-R_{\rm S}/R)^{1/2}=\gamma$.

 %shell of  order of $10^4 (m/10)$ cm. 
 A significant fraction of these annihilation line photons strongly gravitationally  redshifted  can be directly seen by the Earth observer  as  a bump 
 %``a BB-like'' bump   
 located at  $\sim 20$ keV and presumably related to the  representative value of $\gamma\sim 20$. Laurent \& Titarchuk (2009, in preparation) made 
 extensive Monte Carlo simulations of the X-ray spectral formation in the converging flow taking into account photon-electron, photon-photon and pair-electron interactions. These simulations  confirm our expectations
 that  in some cases the emergent spectra of IS and HSS consists of the redshifted annihilation line located at $\sim 20$ keV,  which can be fitted by  ``high temperature BB-like'' profile,  and also the simulated spectra  are extended  to  energies of order of a few MeV [see Fig. \ref{sp_bbody} and \cite{grove98} for details of IS-HSS spectral components].
 % and for the extended power-law tail \cite{grove98}]. 
 %see  to a few MeV as that observationally found 
% by 
%   which lead to  the formation of the annihilation line photons at  the narrow shell there
%(50-100 meters shell for 10 solar mass  BH)
%  for the jet formation in black holes 
%and an extension of the photon spectrum up to a few MeV.  

\subsection{Radio$-$X-ray connection}
 \cite{mfk05}, hereafter MFK05 reported on correlations between radio luminosity and 
X-ray timing (QPO) features in X-ray binary systems containing low magnetic field 
neutron stars and black holes. The MFK05 conclusions 
%Results of \cite{migl05} 
on the radio-QPO 
correlation based on  observations of seven neutron star and one black hole GX~339-4. For 
GX~339-4 they used data only in low-hard state before and after outburst. 
 %\subsection{Instability of diffusive propagation of  disk accretion perturbations and X-ray/radio flares} 
\cite{ts05} based on the analysis of {\it RXTE} data from NS 4U1728-34 confirmed
 a correlation of X-ray and radio emissions with LF QPOs  through all spectral states for this particular NS source. However,  we do not find a real low-hard state in GRS 1915+105 for which photon index $\Gamma_1$ should be about 1.5, as that in GX 339-4 (see ST09),  and therefore we cannot confirm or refute   the radio-QPO  correlation in GRS 1915+105 LHS similar to that found by  MFK05 in GX 339-4.
 
On the other hand we find that LFQPOs  do not correlate with the radio flux while they correlate with the hard component photon index $\Gamma_1$ through IS and HSS 
(see Fig. \ref{radio_QPO_independence} and Fig.  \ref{outburst_index_qpo} respectively). 
Absence of correlation between radio luminosity and QPO can be explained by the different origins of these quantities. While the QPO phenomenon is probably  related to the transition layer oscillations (see \S4.1), it is confirmed by the index-QPO correlation, the  radio emission is presumably originated in the wind or wide open jet.  Furthermore, because the radio flux and iron line EW are strongly correlated 
(see Fig. \ref{outburst_EW_radio}) one can conclude that the iron line is also formed in the wind
[see more on the line formation origin in LT07, \cite{stl09} and \cite{tls09}].   It is also worth noting that the X-ray  flux, and photon index do not correlate with  the radio  flux  (see Fig. \ref{outburst_index_radio_X-ray}).  It  can be explained by the different mechanisms of the energy releases in X-ray and radio.  X-ray radiation  is  presumably formed  in the innermost part of the disk and in the transition layer (or Compton cloud) while the radio emission is presumably formed in the jet or winds  which are probably launched at the outskirts of the disk  [see e.g. \cite{mm99}, \cite{m00},   \cite{mm03} and TSA07].        
%The observational indication  

Using the aforementioned correlations of EW with radio flux  we can suggest that some fraction  of  accretion  flow may  go  to the outflow. Probably the powerful  outflow is  launched  at outer parts of the accretion disk and it  is not by chance that we can see this strong correlation of the iron line EW  with the radio flux. Thus there are two ways for the matter to proceed:
i.	in the outflow if the local mass accretion rate in the disk  exceeds the critical value (which is  proportional to  radius, see TSA07),
ii.	in the disk where  the matter  proceeds  and it ultimately    converges onto BH.  This final stage of the accretion we observe as a saturation of index with mass accretion flow (converging flow signature).
 
Whereas we do not find any correlation between radio and X-ray fluxes during the spectral transition, 
although, probably, we see an indication of X-ray-radio connection during a minor flare event. 
In Figure \ref{radio_appearances} we show the spectral and timing properties of X-ray emission of a typical  minor X-ray/radio flare  As one can see from this Figure (see low panels there) the PDS and energy spectra are different at the peak of the flare from those before and after the flare. Specifically the soft component is more pronounced 
%in the peak flare spectrum 
but  QPO features  are not seen   at the peak of the flare while they are present before and after the flare.  Moreover the  flat part (white noise) of the peak PDS is extended to higher frequencies (break frequency $\nu_b\sim10$ Hz at the peak vs 
 $\nu_b\sim 2$ Hz before and after the flare). 
 
In \S \ref{minor_flare} we reported the results of the  study  the energy dependence of the PDS shape and integrated power variability as a function of
the photon energy.  In Figure \ref{radio_appearances}  we show 
two power spectra for  two energy bands 
2-15 keV (red) and 15-30 keV (blue). 
One can see that PDSs  weakly depend   on the energy band.
We can suggest that the sizes of the photon emission areas $L_{CC}$ related to these two energy bands are also the same. In fact,  \cite{tbw01} argue that  $\nu_{QPO}$ is proportional to the ratio of magneto-acoustic (plasma) velocity  $V_{MA}$ and Compton cloud size   $L_{CC}$ and hence one can conclude that
 %Thus  if $V_{MA}$ is determined by a spectral state only 
 %Then 
 the emission areas  are the same  because  $\nu_{QPO}$ are the same  for these two energy bands.
 
 \cite{le74}, hereafter LE74,  suggest that the thin disk is always  unstable in the inner region when radiation pressure dominates gas pressure.   Using numerical simulations  \cite{l74} found  that the innermost region of a disk  around a BH is secular unstable against clumping of the gas into rings which observational appearances can be seen by the Earth observer as  X-ray-radio flares.  We can speculate that an increase of the soft and hard  components in the observable X-ray spectrum  at the flare peak 
 (compare panels  B2 and A2, C2  of Fig. \ref{radio_appearances}) can be a sign that  the radiation pressure gets to dominate in the inner disk region.   On the other hand the sign of the destruction of some part of the innermost part of the disk, as an effect of the high pressure instability,   should be seen in the PDS.  TSA07 argue that the break frequency $\nu_b$ in PDS  is proportional to the diffusion frequency  $\nu_d=1/t_{visc}\sim {\hat \nu}/R^2$ where $\hat \nu$ is a viscosity,  $t_{visc}$ is a viscous timescale and $R$ is a radial size of the innermost part of the  disk.    Given that $\nu_b$  increases 
at the peak with respect of that before and after the flare  (compare panels  B1 and A1, C1  
of Fig. \ref{radio_appearances}) it can imply that the size $R$ decreases when  $\nu_b$ increases,
i.e. some part of the innermost part of the disk is probably destroyed.

In terms of the diffusion theory,   this disk instability arises in the inner region where the viscous stress $W$ is a decreasing function of the surface density $\Sigma$ and thus an effective diffusion coefficient of the nonlinear equation for  $\Sigma$  becomes {\it negative} there [see Eqs. 4-5 in LE74]. On the other hand Makeev \& Titarchuk (2009, in preparation),  hereafter MT09,  obtain this disk instability  as a solution  the linear diffusion equation for  $\Sigma$ and they do not specify any (ad hoc) assumption about the nature of the disk viscosity (cf. LE74).
 They just assume   the power-law  viscosity distribution over the disk and they use  the TLM98 angular velocity distribution  in the transition layer (TL). 
 %\cite{tsa07}, hereafter TSA07,  push forward an idea
%in their recent work followed the route
%outlined by Lyubarskii (1997), extending his development by the idea 
%that timing  properties of the X-ray emission
%can  result from diffusive propagation of the "driving" perturbations in a bounded medium of the
%accretion flow (TL or accretion disk). These driving perturbations can be initiated at any radius in the disk by, for example,
%Rayleigh-Taylor local instabilities (common for systems with inhomogeneous vertical density structure). The other possible sources of  fluctuations include 
%gravity waves, 
%local  fluctuations of the eccentricities of particle
%orbits resulting in perturbations of the effective viscosity, variations associated with magnetic stresses which may be variable as a result of small$-$scale dynamo.
% \cite{tm09}, 

MT09  study the TSA07 model of the PDS formation   and they find that this model  predicts the existence of the two distinct zones within the TL of the black hole or neutron star, with oppositely different types of  diffusion of perturbation  taking place in each zone. A simple fact of the change of sign of the angular velocity derivative in the linear diffusion equation for $\Sigma$ at the critical radius $R_{max}$ results  in  a {\it negative} diffusion coefficient  of the equation
 in the interval  $R_{in}<r<R_{max}$.   Moreover  the change of sign of the diffusion coefficient at  $R_{max}$  leads  to a turnover of the angular momentum transfer, changing it towards the central body of the accreting system, instead of being pushed outwards as in the Keplerian disk.   
One of the considered scenarios implies  an unstable diffusion of the perturbations in the TL inner zone  which might be an indication of the development of a X-ray flare followed by a radio flare. 
The absence  of any quasiperiodic oscillations in the TL  and raising  $\nu_b$ are  possible indications  of this instability.

\section{Conclusions \label{summary}} 
We concentrate our efforts on the study of the
correlation between the spectral index and 
%characteristic sub-second 
%variability frequencies  and
the  accretion disk luminosity. 
%More fundamentally, 
We argue that the shape of the correlation
pattern can contain the direct BH signature. 
Namely, we show both observationally and theoretically 
that the index saturates with mass accretion rate  which is a signature of a converging 
flow. This index saturation effect  can exist only in  BH sources.
Also this  correlation 
pattern carries the most direct
information on the BH mass and the distance to the source (see ST09).

We compiled the state transition data from GRS 1915+105
collected with the \textit{RXTE} mission. We examined the correlation between the photon 
index of the Comptonized spectral component, its normalization and the
QPO frequency (see Figs.~\ref{outburst_97_rise}-\ref{outburst_05-06_decay},
\ref{outburst_index_norm}-\ref{outburst_index_qpo}).

The spectral data of GRS 1915+105 are well fitted by two  (soft and hard)  BMC components for most of analyzed IS and HSS spectra (see Fig. \ref{two_BMC}) 
while LHS spectra essentially require only one BMC component.
% (see panel S1 in Fig. \ref{spec_evol_4Sm}).
 In addition to two BMC components  8  IS-HSS spectra require   an extra component which can be fitted by `` high temperature BB-like"  profile.  We suggest this ``BB'' component is probably a signature of the redshifted annihilation line formed in the very narrow shell near a BH horizon due to high photon compactness  taking place during  intermediate  and high-soft states  (see Fig. \ref{sp_bbody} and Table 7).

A remarkable result of our study is that  the index - normalization (mass accretion rate)
 correlation seen  in GRS 1\-9\-1\-5+105 is predicted  by the theory of the 
converging flow.
% now is confirmed by the observations of 
We  demonstrate that a strong index saturation vs disk flux  seen in the index-disk flux  
correlation   (see Fig.~\ref{outburst_index_norm}) is an observational signature of the presence of the converging  flow, which should only exist  the BH sources.
 In other words, {\it this index saturation effect provides a robust observational 
evidence for the presence of black hole in GRS~1915+105}.

%We  find particular features of power spectra (PDS) when QPOs occur at X-ray/radio flare events. 
%In fact,  QPOs are only present, in the corresponding PDSs,  before  and after X-ray/radio flare 
%However  we do not find any sign of QPOs at the peak of the flare. The related  PDS is featureless.
% with no sign of QPO
%(see Fig.~\ref{radio_appearances}).
We also  find  a tight positive correlation of QPO frequencies with the index  (see Fig. \ref{outburst_index_qpo}) and consequently that with the disk flux.

Our comprehensive analysis of  X-ray and radio emissions in GRS 1915+105 shows that   QPOs are seen independently of radio activity of the source during the spectral transition from low-hard to high-soft state.  Specifically these QPO features have been  detected
at any level of   the radio flux and even when the radio emission is at the noise level
% is very low  and very high 
(see left and right panels in Fig.~\ref{radio_QPO_independence} correspondingly). 
We also do not find any correlation between X-ray and radio fluxes  and X-ray power-law index.
(see Fig. \ref{outburst_index_radio_X-ray}). 
However,  we establish  a  strong correlation between equivalent width of iron line and radio flux in binary GRS~1915+105 
(see Fig. \ref{outburst_EW_radio}).

%In Figures \ref{outburst_index_norm} and  \ref{outburst_index_qpo} we present the index-BMC normalization  and index-QPO correlations inferred 
%from \textit{RXTE}  observations of GRS 1915+105. 
 %ST09 argue that these correlations can be used for BH mass and BH source distance determinations if the scalable (self-similar) patterns of 
% these correlations can be found for  a source for which BH mass and distance is known. One of the goals for our future study  of Black Hole Candidates  
 %is to find such a source for which the correlation patterns are scalable with those in GRS 1915+105.

We are grateful to the referee whose constructive suggestions help us to  improve the paper quality.
We  acknowledge productive discussion with Nikolai Shaposhnikov and  Ada Paizis and we also would like to thank Guy Pooley who 
kindly provided us  {\it Ryle Radio Telescope} data. 

\appendix
\section{The light aberration effect}
For a given electron velocity  ${\bf v}$ moving inward  along radius ${\bf v} =-v{\bf e_r }$ 
 the directions of the ray in  the observed laboratory  and comoving electron  frames $\bf n_0$ and $\bf n$ are related to aberration  formula [see e.g. \cite{rl79}]
\begin{equation}
\cos\theta=\frac{v/c-\cos\theta_0}{(v/c)\cos\theta_0 -1}.
\label{aberration}
\end{equation}
where $\gamma=[1-(v/c)^2]^{-1/2}$ is Lorentz factor, $\cos\theta_0=-(\bf n_0\cdot e_r$) and 
$\cos\theta=-(\bf n\cdot e_r$) are   for  the laboratory  (zero) and electron rest frames respectively. 
%Note the laboratory  frame is an inertial frame of reference, the unadorned frame,  
%move with velocity  $\bf v$ as seen  by the observer  in another frame of reference, the zero frame.

For $\gamma\gg 1$,  we can write that $1-v/c=1/(2\gamma^2)$ and thus we can rewrite Eq.  
(\ref{aberration}) as follows
\begin{equation}
\cos\theta\sim\frac{-2\sin^2(\theta_0/2)+1/(2\gamma^2)}{2\sin^2(\theta_0/2)+\cos\theta_0/(2\gamma^2)}.
\label{aberration_m}
\end{equation}
If $\theta_0\sim b/\gamma$ then for $b>1$ 
\begin{equation}
\cos\theta\sim \frac{1-b^2}{1+b^2}<0
\label{aberration_mm}
\end{equation}
i.e.  ray is directed against the electron direction. In fact, $\cos\theta\sim-1$  for  {\it all} 
$\theta_0\gg1/\gamma$. On the other hand 
\begin{equation}
\cos\theta=-({\bf n\cdot e_r})>0
\label{cos_positive}
\end{equation}
if  $b<1$.

\newpage
\bigskip
\begin{deluxetable}{cccccccccccc}
\rotate
\tablewidth{0in}
\tabletypesize{\scriptsize}
%  \begin{center}
    \tablecaption{Best-fit parameters of spectral analysis of PCA and HEXTE
observation of GRS~1915+105 in 3-150~keV energy range during rise 1997 transition$^{\dagger}$.
% Errors (put in parenthesis) correspond to 1$\sigma$ level.}
Parameter errors (put in parenthesis) correspond to 1$\sigma$ confidence level.}
%\vspace{1em}
    \renewcommand{\arraystretch}{1.2}
%    \begin{tabular}[h]
%      \hline
\tablehead{Observational& MJD   & $\alpha_1=\Gamma_1-1$& log($A_1$)$^{\dagger\dagger}$& $N_{bmc1}$, 
 & $\alpha_2=\Gamma_2-1$ &log($A_2$)$^{\dagger\dagger}$&   $N_{bmc2}$,  & EW,& E$_{laor}$ & Flux$^{\dagger\dagger\dagger}$& $\chi^2_{red}$ (d.o.f.)\\
ID               & day   &             &           &$L_{39}/d^2_{10}$&  &    &$L_{39}/d^2_{10}$  &  eV        &keV &  &}
 \startdata
20402-01-11-00&50462.06&0.867(9)& 0.106(4) &0.115(3)&  1.84(7)  &  2.0      &  0.009(3) &-&- &  13.59 &1.09 (73)\\
20402-01-12-00&50471.06&0.844(7)& 0.102(3) &0.1105(2)&  1.9(2)   &  2.0      &  0.009(2) &-&-    &  13.28 &1.06 (73)\\
20402-01-13-00&50477.87&0.92(2)& 0.030(6) &0.116(1)&   1.9(3)   & 2.0       &  0.002(5) &-&-    &  13.12 &1.01 (73)\\
20402-01-16-00&50501.88&0.83(1)& 0.100(6) &0.1014(5)&   2.0(4)   & 2.0       &  0.008(2) &-&-    &  12.01 &1.07 (73)\\
20402-01-19-00&50517.04&0.62(1)& 0.137(2) &0.101(1)&   1.8(2)   & 2.0       &  0.002(1) &-&-    &  12.16 &1.02 (73)\\
20402-01-20-00&50524.92&0.85(1)& 0.077(8) &0.092(2)&   1.9(3)   & 2.0       &  0.01(2)  &-&-    &  11.51 &1.05 (73)\\
20402-01-21-00&50533.83&0.94(1)& 0.050(3) &0.100(1)&   2.0(4)   & 2.0       &  0.009(5) &-&-    &  11.34 &1.09 (73)\\
20402-01-21-01&50534.90&0.95(1)& 0.025(3) &0.101(1)&   2.0(2)   & 2.0       &  0.009(2) &-&-    &  11.30 &1.06 (73)\\
20402-01-23-00&50548.45&0.91(3)&-0.082(8) &0.112(2)&   2.0(1)   & 2.0       &  0.01(3)  &-&-    &  11.55 &1.09 (73)\\
20402-01-24-00&50561.13&0.72(3)& 0.095(4) &0.116(5)&   2.0(3)   & 2.0       &  0.01(4)  &-&-    &  14.27 &1.08 (73)\\
20402-01-28-00&50586.68&1.59(2)&-0.178(8) &0.2001(2)&2.34(3)&2.0&  0.109(1)&- &-& 18.59 &1.07 (73)\\
20402-01-29-00&50589.49&0.95(2)& 0.033(3) &0.159(1)&  2.0(2)    & 2.0       &  0.01(3)  &-&-   &  18.00&1.07 (73) \\
20402-01-30-01&50596.20&1.59(1)& 0.32(2)   &0.200(8)  &2.3(1)&2.0  & 0.110(2)&- &-& 20.45 &1.03  (73)\\
20402-01-31-00&50602.53&1.79(2)& 0.07(4)   &0.223 (5) &2.6(1)&0.09(6) & 0.118(5)&              - &-& 23.56 &1.01 (73)\\
20402-01-31-02&50604.61&1.80(5)&-0.11(8)   &0.218(7)  &2.6(1) &0.8(1)& 0.127(6)&                - &-& 23.67   &1.08 (73)\\
20402-01-33-00&50617.54&1.99(6)& 0.13(6)   &0.22(4)    &3.2(2) &0.18(1)& 0.24(1) &               - &-& 30.21  &1.01 (73)\\
20402-01-34-01&50621.80&1.92(4)& 0.49(9)   &0.22(1)    &3.1(1) &-0.18(2) & 0.17(1) &             - &-& 27.18&1.07 (73)\\
20402-01-35-00&50636.62&1.85(6)& 0.030(3)  &0.24(4)    &3.10(9)&2.0&0.21(5)&  - &-& 31.54 &0.97 (73)\\
20402-01-36-00&50639.62&1.99(9)&-0.40(1)   &0.20(1)    &3.2 (1) &2.0&0.256(5)&  - &-& 32.24&0.97 (73)\\
20402-01-37-01&50641.48&1.90(9)&-0.26(9)   &0.245(9)  &3.09(9)&2.0   & 0.24(1) & - &-& 34.38  &1.03 (73)\\
20402-01-38-01&50649.42&1.9(1)&-1.66(8)     &0.370(9)  &3.20(5)&2.0   & 0.30(1) & 63 (8)  &6.1$\pm$0.7 & 38.59  &1.10 (73)\\
20402-01-39-02&50658.51&2.0(1)&-0.58(7)     &0.26(2)    &3.2(3) &2.0  & 0.303(2)&61(10)  &6.1$\pm$1.5 & 38.15 &1.12 (73)\\
20402-01-41-01&50679.30&2.0(1) &-1.82(9)    &0.417(6)  &3.15(5)&0.7(1)& 0.422(8)&         78(10)  &6.1$\pm$1.0 & 46.47 &1.10  (73)  \\
20402-01-41-02&50679.37&2.00(9)&-1.40(8)   &0.459(2)  &3.20(1)&2.0   & 0.429(1)&82(10)  &6.09$\pm$0.8& 49.90 &1.08 (73)\\
20402-01-43-00&50688.24&2.00(5)&-0.19(6)   &0.342(5)  &3.20(7)&2.0   & 0.236(3)&124(14) &6.12$\pm$0.6& 41.87 &1.01 (73)\\
20402-01-45-00&50698.65&1.90(1)&-0.22(9)   &0.231(2)  &3.20(2)&0.47(5)      & 0.236(2)&         151(13) &5.78$\pm$1.4& 31.22 &0.96 (73)\\
      \enddata
 %     \hline
%      \end{tabular}
    \label{tab:table}
%  \end{center}
$^\dagger$ The spectral model of the continuum is  $wabs * (bmc + bmc*highecut + laor)$,
$^{\dagger\dagger}$ when parameter $log(A)>1.0$, this parameter is fixed to a value 2.0 (see comments in the text), 
$^{\dagger\dagger\dagger}$spectral flux in the 3 -- 150 energy range in units of $\times 10^{-9}$ ergs/s/cm$^2$.
%* this observation was fitted with systematic error $>$1\%
\end{deluxetable}

%middle 1997
\begin{deluxetable}{cccccccccccc}
\rotate
\tablewidth{0in}
\tabletypesize{\scriptsize}
%  \begin{center}
    \tablecaption{Best-fit parameters of spectral analysis of PCA and HEXTE
observation of GRS~1915+105 in 3-150~keV energy range during middle 1997 transition.
% Errors (put in parenthesis) are given at the 99\% confidence level}
Parameter errors (put in parenthesis) correspond to 1$\sigma$ confidence level.}
     \renewcommand{\arraystretch}{1.2}
\tablehead{Observational& MJD   & $\alpha_1=\Gamma_1-1$& log($A_1$)$^\dagger$& $N_{bmc1}$, 
 & $\alpha_2=\Gamma_2-1$ &log($A_2$)$^\dagger$&   $N_{bmc2}$,  & EW,& E$_{laor}$ & Flux$^{\dagger\dagger}$& $\chi^2_{red}$ (d.o.f.)\\
ID               & day   &             &           &$L_{39}/d^2_{10}$&  &    &$L_{39}/d^2_{10}$  &  eV        &keV &  &}
 \startdata
20402-01-45-02& 50696.21&  1.84(1)& 0.35(9)&0.33(1)&2.98(2)&0.56(1)&0.379(2)  & 133(5)&6.31$\pm$0.08 & 39.45 &1.05 (73)\\
20402-01-45-03& 50700.25&  1.96(4)& -0.1(2)&0.23(4)&3.19(6)&0.49(2)&0.245(8)  & 192(8)&6.10$\pm$0.08 &33.49 &1.01 (73)\\
20402-01-46-00& 50703.41&  2.00(7)& 0.02(1)&0.25(2)&3.1(2)  &0.8(4)  &0.13(2)    & 87(10)&6.25$\pm$0.08 & 26.96 &1.09 (73)\\
20402-01-48-00& 50720.59&  2.00(7)& 0.77(4)&0.240(6)&3.09(3)&0.15(2)&0.24(1)& 164(13)&6.46$\pm$0.05& 39.44 &1.08 (73)\\
20402-01-50-00& 50735.54&  1.74(7)& 0.16(5) &0.222(8)&3.2(2)&2.0&0.323(9)&125(19)&6.48$\pm$0.01&20.69 &1.02 (73)\\
20402-01-50-01& 50737.40&  1.720(2)& 2.0&0.236(1)&  1.9(4) & 2.0      & 0.02(2) &102(16)&6.47$\pm$0.04&19.19 &1.02 (73)\\
20402-01-51-00& 50743.29&  1.704(2)& 1.9(6)&0.226(1)&2.0(3) & 2.0      & 0.04(5) &102(13)&6.47$\pm$0.03&20.56 &1.09 (73)\\%%
20402-01-52-01& 50751.68&  2.00(4)& 0.15(9)&0.240(6)&3.2(1)&2.0&0.271(6) & 348(19)&6.31$\pm$0.01&34.60 &1.02 (73)\\
20402-01-52-02& 50751.75&  2.00(8)& 0.06(1) &0.320(2)&3.2(1)&2.0&0.482(5)& 342(20)&6.44$\pm$0.01&49.68 &1.04 (73)\\
20402-01-53-00& 50752.01&  2.00(2)& -0.46(1)&0.32(1)&3.00(7)&2.0&0.43(1)  & 315(10)&6.10$\pm$0.01&48.71 &0.99 (73)\\
20402-01-53-02& 50756.41&  2.00(2)& -0.45(3)&0.39(1)&3.20(1)&2.0&0.32(2)  & 53(12)&6.18$\pm$0.07&43.76 &1.01 (73)\\
20402-01-54-00& 50763.20&  1.99(8)& 2.0 &0.412(6)&3.20(3)&2.0&0.369(7)&58(16)&6.19$\pm$0.08&45.09 &1.02 (73)\\
20402-01-55-00& 50769.22&  0.8(3) & 0.08(6)      & 0.01(6) &3.15(3)&0.322(1)&0.426(4)  &101(13)&6.25$\pm$0.09&35.88&1.02 (73)\\
      \enddata
%%      \hline
%%      \end{tabular}
    \label{tab:table}
%%  \end{center}
%%\end{table}
$^\dagger$
%This parameter is fixed,
When parameter $log(A)>1.0$, this parameter is fixed to a value 2.0 (see comments in the text), 
$^{\dagger\dagger}$spectral flux in the 3
-- 150 energy range in units of $\times 10^{-9}$ ergs/s/cm$^2$.
%, * this observation was fitted with systematic error $>$1\%
\end{deluxetable}

%decay 1998
\begin{deluxetable}{cccccccccccc}
\rotate
\tablewidth{0in}
\tabletypesize{\scriptsize}
%  \begin{center}AAAAAAAAAAAAAAAAAAAA
    \tablecaption{Best-fit parameters of spectral analysis of PCA and HEXTE
observation of GRS~1915+105 in 3-150~keV energy range during decay 1997 transition.
 %Errors (put in parenthesis) are given at the 99\% confidence level}
Parameter errors (put in parenthesis) correspond to 1$\sigma$ confidence level.}
    \renewcommand{\arraystretch}{1.2}
\tablehead{Observational& MJD   & $\alpha_1=$& log($A_1$)$^\dagger$& $N_{bmc1}$, 
 & $\alpha_2=$ &log($A_2$)$^\dagger$&   $N_{bmc2}$,  & EW,& E$_{laor}$ & Flux$^{\dagger\dagger}$& $\chi^2_{red}$ (d.o.f.)\\
ID               & day   &  $\Gamma_1-1$ &           &$L_{39}/d^2_{10}$&$\Gamma_2-1$  &    &$L_{39}/d^2_{10}$  &  eV        &keV &  &}
 \startdata
30184-01-01-00& 50908.00& 1.81(2)&0.32(2) &0.260(1) &  2.0(3)      &   2.0     & 0.009(1)  &284(18)&5.75$\pm$1.2 &  24.90 & 1.09 (73)\\
30703-01-14-00& 50909.87& 1.74(1)&0.55(2) &0.236(1) &  2.0(5)      &   2.0     & 0.007(2)  &317(20)&6.39$\pm$0.05&  22.85 & 1.30 (73)\\
30402-01-09-00*& 50912.88& 1.50(2)&-0.32(9)  &0.20(3)     &2.09(5) &2.0& 0.11(1) &290(18)&5.7$\pm$0.3& 23.89 & 1.04 (71)\\
30402-01-10-00& 50914.39& 1.50(1)&-0.32(8)  &0.1949(8) &2.09(2) &2.0& 0.111(9)&411(30)&5.7$\pm$0.4& 22.92 & 1.02 (73)\\
30402-01-11-00& 50923.26& 2.00(3)&1.0(2) &0.24(1)     &3.2(1)   &2.0& 0.23(3) &308(20)&5.6$\pm$0.2 &  31.28 & 1.23 (73)\\
30703-01-15-00& 50925.88& 1.79(5)&0.37(6) &0.22(3)     &2.1(1)  &2.0& 0.10(1) &316(10)&6.4$\pm$0.2& 23.79 & 1.01 (73)\\
30703-01-16-00& 50931.67& 1.79(4)&0.37(6) &0.21(3)   &2.10(9)&2.0   & 0.09(2) &326(8) &6.4$\pm$0.3& 23.52 & 1.07 (73)\\
30402-01-12-01& 50945.01& 1.69(2)&0.42(3) &0.206(4)   &2.10(8)&-0.5(1)      &0.094(2)&290(10)&5.7$\pm$0.5& 22.17  & 1.06 (73)\\
30182-01-02-00& 51003.21& 1.801(2)&2.0 &0.223(2) &    2.0(3)  &  2.0 &  0.009(1)     &434(11)&6.41$\pm$0.01& 26.11 & 0.99 (73)\\
30182-01-04-01& 51006.21& 1.0(4)&-1.953(8)  &0.169(8)  &2.49(6)& 2.0& 0.125(2)&- &-&24.23 & 1.03 (73) \\
30402-01-17-00& 51067.62& 1.520(3)&0.6(1)  &0.200(1) &  2.1(3)&  2.0       &  0.009(2)  &- &-& 21.15 & 0.98 (73)\\
30703-01-33-00& 51071.90& 1.141(7)&0.19(1)&0.172(1)&   2.0(4)&  2.0       &  0.002(3)  &- &-& 19.52 & 1.07 (73)\\
30703-01-35-00& 51081.81& 0.88(3)&0.18(4) &0.152(2) &  2.1(2) & 2.0       &  0.009(1)  &- &-& 18.18 & 1.02 (73)\\           
      \enddata
    \label{tab:table}
$^\dagger$
When parameter $log(A)>1.0$, this parameter is fixed to a value 2.0 (see comments in the text),     
%$^\dagger$This parameter is fixed, 
$^{\dagger\dagger}$spectral flux in the 3
-- 150 energy range in units of $\times 10^{-9}$ ergs/s/cm$^2$, 
*the data are fitted with {\it wabs*(bmc+bmc*highecut+laor+bbody)} model, see values of the best-fit BB color temperature and EW  in Table 7.
\end{deluxetable}

%\end{deluxetable}  

\newpage

% rise 2005

\begin{deluxetable}{cccccccccccc}
\rotate
\tablewidth{0in}
\tabletypesize{\scriptsize}
%  \begin{center}
    \tablecaption{Best-fit parameters of spectral analysis of PCA and HEXTE
observation of GRS~1915+105 in 3-150~keV energy range during rise 2005 transition.
%Errors (put in parenthesis)  are given at the 99\% confidence level}
Parameter errors (put in parenthesis) correspond to 1$\sigma$ confidence level.}
    \renewcommand{\arraystretch}{1.2}
\tablehead{Observational& MJD   & $\alpha_1=\Gamma_1-1$& log($A_1$)$^\dagger$& $N_{bmc1}$, 
 & $\alpha_2=\Gamma_2-1$ &log($A_2$)$^\dagger$&   $N_{bmc2}$,  & EW,& E$_{laor}$ & Flux$^{\dagger\dagger}$& $\chi^2_{red}$ (d.o.f.)\\
ID               & day   &             &           &$L_{39}/d^2_{10}$&  &    &$L_{39}/d^2_{10}$  &  eV        &keV &  &}
 \startdata
80701-01-48-00&53382.39&1.78(5)&-0.96(1)& 0.125(1)& 2.1(4)&  2.0  & 0.009(2) &- &-& 6.25 &1.05 (73)\\
90701-01-46-00&53400.35&2.00(1)&-0.21(2)& 0.292(4) &2.90(1)& 1.1(4)& 0.335(1) &91(10)&6.35$\pm$0.01& 29.94&0.96 (73)\\
80701-01-37-00&53416.34&2.00(4)&2.0& 0.164(1)&2.92(3)& 2.0&0.157(4)&594(10)&5.5$\pm$0.2&24.03&1.03 (73)\\
90701-01-49-00&53422.30&2.00(1)&0.26(9)& 0.152(4)  &3.00(2)& 0.75(5)&0.158(4)&626(15)&6.56$\pm$0.02& 22.30&1.01 (73)\\
90105-05-03-00&53442.98&1.9(1)&2.0& 0.204 (3)   &2.92(4)& 2.0& 0.1574(6)&435(15)&6.45$\pm$0.01&24.37&1.06 (73)\\
%90105-05-03-01*&53443.03&2.00(9)&2.0& 0.242(3)  &2.98(7)& 2.0& 0.1411(6)& 24.37&1.83 (73)\\
%90105-05-03-03*&53443.24&1.950(3)&2.0& 0.227(2)&2.80(6)& 2.0& 0.1176(6)& 24.03&1.24 (73)\\
90105-05-03-04&53444.01&1.981(6)&2.0& 0.191(3)&2.7(0.1)& 0.880(2)& 0.111(2)     &402(12)&6.45$\pm$0.01& 23.08&1.09 (73)\\
90105-05-03-05*&53444.08&1.982(4)&2.0& 0.187(1)&2.79(6)& 0.88(5)  & 0.112(1)   &378(10)&6.46$\pm$0.01& 22.67&0.96 (71)\\
91701-01-04-00*&53456.28&1.981(5)&2.0& 0.185(2)&2.79(9)& 0.884(1)& 0.110(1)   &256(15)&6.33$\pm$0.01& 22.69&0.97 (71)\\
90105-07-01-00&53472.92&1.980(6)&2.0& 0.144(1)&2.85(6)& 0.251(1)& 0.112(1)   &55.6(10)&6.57$\pm$0.01& 22.07&0.99 (73)\\
90105-07-02-00&53473.05&1.981(5)&1.0(2)& 0.147(5)&2.80(6)& 0.272(3)& 0.133(3)        &75.3(6)&6.55$\pm$0.07& 22.36&0.99 (73)\\
90105-07-03-00&53473.97&1.98(1)&2.0& 0.147(2)  &2.80(8)& 2.0&0.135(8)   &210(8)&6.39$\pm$0.01& 22.61&0.96 (73)\\
91701-01-08-00&53486.12&1.96(6)&2.0& 0.16 (3) &3.1(2)& -0.02(3) & 0.38(3)       &132(11)&6.13$\pm$0.04&40.68&1.01 (73)\\
91102-01-01-00&53488.72&1.60(5)&-0.47(7)& 0.123(3)&3.0(1)& -0.0030(5)& 0.4705(5)    &74(12)&6.32$\pm$0.08& 38.69&1.08 (73)\\
91412-01-01-00&53500.32&1.99(2)&-1.10(8)& 0.261(1)&3.10(1)& 0.235(9)&0.3400(9)     &50(16)&6.19$\pm$0.09& 39.35&1.02 (73)\\
91701-01-10-00&53501.06&2.0(1)&-1.11(9)& 0.338(2)  &3.20(2)& 0.34(1) &0.456(1)        &- &-& 51.99&1.11 (73)\\
91701-01-10-01&53501.12&1.9(1)&-1.06 (8)& 0.210(4)  &3.20(2)&2.0 &0.433(4)      &68(9)&6.10$\pm$0.04& 45.30&1.39 (73)\\
90105-08-02-00&53503.87&1.90(3)&-0.86 (6)& 0.358(1)&3.20(1)&0.802(8)& 0.4901(6)    &104(10)&6.3$\pm$0.1& 57.53&1.09 (73)\\
90105-08-03-00&53504.71&2.00(1)&-0.86(7)& 0.242(2) &3.20(1)& 0.8(2)    &0.278(1)       &180(12)&6.11$\pm$0.01& 37.13&1.02 (73)\\
91701-01-11-00&53508.06&1.70(3)&-1.7  (9)&  0.343(2)&3.10(1)& 2.0 &0.451(1)    &69(10)&6.13$\pm$0.04& 56.21&1.04 (73)\\
91701-01-12-00&53515.07&0.8(1)  &0.3(1)       &  0.001 (5) &3.197(7)&2.0  &0.3480(1)&90(10) &6.4$\pm$0.1&24.18&1.09 (73)\\
91701-01-13-00&53520.05&0.8(1)  & 0.9(4)      &  0.05 (6) &3.200(3)& 2.0 &0.430(1)& 120(10) &6.4$\pm$0.1&37.02&1.10 (73)\\
91701-01-17-00&53547.93&0.7(1)  & 2.0&  0.06 (2) &2.75(3)&-1.106(1)&0.193(1)    &95(10) &7.1$\pm$0.2& 11.24&0.99 (73)\\
91701-01-19-00&53562.92&0.7(1)  & -0.9(5)     &  0.09 (5) &3.20(1)& 0.256(7)& 0.4633(7)&100(10)&7.0$\pm$0.1& 23.69&1.02 (73)\\
91701-01-20-00&53570.78&0.7(2)  & -0.9(3)     &  0.05 (3) &2.57(1)&-0.537(8)& 0.151(1)&125(10) &7.0$\pm$0.3& 69.91&1.18 (73)\\
%91701-01-23-00*&53588.83&0.7(1)& -0.54(2)   &  0.112 (5) &2.7(1)&-0.624(1)& 0.124(1)  5.76&1.2 (73)\\
91701-01-24-00&53600.89&0.7(1)  & -0.95(5)    &  0.123 (6) &2.75(9)&-0.627(1)& 0.124(1)   &100(10) &7.0$\pm$0.3& 5.90&1.24 (73)\\
91701-01-25-00&53606.84&0.7(2)  & -0.7(3)     &  0.09 (5) &2.75(7)&-0.626(1)& 0.132(1)   &100(10) &7.1$\pm$0.2& 6.22&1.25 (73)\\
90105-04-01-00&53640.38&0.7(1)  & 0.4(1)      &  0.04 (5) &2.16(2)&-0.530(4)& 0.1335(4)&127(5)&6.40$\pm$0.09& 6.64&1.35 (73)\\
90105-04-03-00&53641.43&1.8(6)    &0.70(4) &0.215(4)&3.00(1)& 1.1(2) & 0.282(3) &114(12)&6.11$\pm$0.07& 33.96&1.08 (73)\\
90105-04-03-01&53641.51&1.8(7)    &0.48(4) &0.215(4)&3.00(1)& 1.1(1) & 0.284(3) &108(11)&6.32$\pm$0.01& 32.26&1.01 (73)\\
      \enddata
    \label{tab:table}
$^\dagger$
When parameter $log(A)>1.0$, this parameter is fixed to a value 2.0 (see comments in the text),     
%$^\dagger$this parameter is fixed, 
$^{\dagger\dagger}$spectral flux in the 3 -- 150 energy range in units of $\times 10^{-9}$ ergs/s/cm$^2$, 
* these data are fitted with {\it wabs*(bmc*highecut+bmc+laor+bbody)} model, see   values of the best-fit  parameters in Table 7.
%, * this observation was fitted with systematic error $>$1\%
\end{deluxetable}

% 2005 - middle

\begin{deluxetable}{ccccccccccccc}
\rotate
\tablewidth{0in}
\tabletypesize{\scriptsize}
%  \begin{center}
    \tablecaption{Best-fit parameters of spectral analysis of PCA and HEXTE
observation of GRS~1915+105 in 3-150~keV energy range during middle 2005 transition.
%Errors (put in parenthesis)  are given at the 99\% confidence level}
Parameter errors (put in parenthesis) correspond to 1$\sigma$ confidence level.}
    \renewcommand{\arraystretch}{1.2}
\tablehead{Observational& MJD   & $\alpha_1=\Gamma_1-1$& log($A_1$)$^\dagger$& $N_{bmc1}$, 
 & $\alpha_2=\Gamma_2-1$ &log($A_2$)$^\dagger$&   $N_{bmc2}$,  & EW,& E$_{laor}$ & Flux$^{\dagger\dagger}$& $\chi^2_{red}$ (d.o.f.)\\
ID               & day   &             &           &$L_{39}/d^2_{10}$&  &    &$L_{39}/d^2_{10}$  &  eV        &keV &  &}
 \startdata
91701-01-31-00& 53646.76&  2.00(2)& 2.0 & 0.17(1)  &3.0(1)& 2.0& 0.176(8)   &- &-& 22.77&1.03 (73)\\
91701-01-33-01& 53659.98&  1.60(1)& 2.0 & 0.087(3)&2.90(6)& 2.0 & 0.143(2) &110(13)&6.35$\pm$0.07& 25.55&1.03 (73)\\
91701-01-33-00& 53661.70&  1.99(1)& 2.0 & 0.112(2)&2.78(1)& 2.0 &0.403(1)&92(8)&6.30$\pm$0.01& 36.88&0.99 (73)\\
91701-01-34-00& 53669.69&  1.42(1)& 2.0 & 0.055(1)&2.51(1)& 2.0 & 0.189(1) &135(10)&6.32$\pm$0.01& 21.21&1.08 (73)\\
91701-01-34-01& 53669.75&  1.7(2)& -0.57(7)& 0.084(2)  &1.98(4)& 2.0 & 0.185(1)    &- &-&  21.14&1.04 (73)\\
91701-01-35-00& 53674.59&  1.50(5)&-0.55(2)& 0.116(2) &3.00(2)&  2.0& 0.3127(6)  &- &-&  23.31&1.01 (73)\\
90105-06-03-01& 53694.90&  1.70(3)& -1.4(2)& 0.345(1)&3.200(1)&  2.0& 0.552(1)&- &-&  62.96&1.02 (73)\\
90105-06-03-00& 53695.03&  1.89(2)& -0.56(7)& 0.261(4)&3.20(1)&  2.0& 0.417(5)   &- &-&  49.38&1.01 (73)\\
90105-06-03-02& 53695.30&  1.7(1)& -0.70(8)& 0.282(6) &3.18(3)&  2.0& 0.389(5)    &- &-&  48.40&1.07 (73)\\
91701-01-38-00& 53696.56&  1.80(8)& -1.1(1) &0.35(1)  &3.12(2)&  0.235(1)& 0.479(1)     &- &-&  51.76&0.99 (73)\\
91701-01-38-01& 53696.63&  2.07(3)& 0.68(8)& 0.685(4)  &2.7(2)&  2.0 & 0.290(3)    &- &-&  57.06&0.98 (73)\\
91701-01-39-00& 53703.57&  1.70(2)& -0.22(9)& 0.159(3) &2.90(1)& 2.0 & 0.273(4)  &- &-&  31.30&1.11 (73)\\
91701-01-39-01& 53703.63&  1.734(7)& 0.35(2)& 0.170(1)  &2.98(3)&  0.471(1)& 0.163(1)  & - &-& 27.10&1.05 (73)\\
92092-01-01-01& 53704.94&  1.74(3)& 0.05(1)& 0.195(3)  &2.9(1)&  2.0& 0.256(2)     &- &-&  28.45&1.05 (73)\\
92092-02-01-01& 53706.00&  1.91(2)& 2.0 & 0.116(3) &2.9(1)&  -0.363(1)& 0.229(1)  &- &-&  22.75&0.95 (73)\\
92092-03-01-01& 53707.68&  1.7(3) & -1.6(1) & 0.36(2)  &3.00(6)&  0.470(1)&0.49(1)        &- &-&  56.72&1.08 (73)\\
91701-01-39-01& 53711.55&  1.8(1)&-0.33(5)& 0.23(2)  &2.9(1)&  2.0 & 0.26(2)        &- &-&  34.48&0.96 (73)\\
91701-01-41-01& 53718.56&  1.77(2)& 0.85(3)&0.184(1) & 2.1(5) &  2.0 &  0.009(1)  &- &-&  19.95&1.06 (73)\\
91701-01-42-00& 53723.47&  1.701(5)& 0.8(1)  & 0.111(2)  &2.90(6)&  2.0 & 0.110(1)&- &-&  20.57&1.13 (73)\\
91701-01-42-01*& 53723.54&  1.91(1)& 2.0 & 0.129(3) &2.70(6)&  2.0 & 0.1644(5)&- &-&  20.57&1.05 (71)\\
91701-01-43-00*& 53730.41&  1.91(4)& 2.0 & 0.15(2) &2.7(2)&  2.0 & 0.14(9)   &- &-&  20.29&0.91 (71)\\
      \enddata
    \label{tab:table}
$^\dagger$
When parameter $log(A)>1.0$, this parameter is fixed to a value 2.0 (see comments in the text),     
%$^\dagger$this parameter is fixed, 
$^{\dagger\dagger}$spectral flux in the 3 -- 150 energy range in units of $\times 10^{-9}$ ergs/s/cm$^2$,
* these data are fitted with {\it wabs*(bmc*highecut+bmc+laor+bbody)} model, see   values of the best-fit  parameters in Table 7.
\end{deluxetable}

%decay 2005-2006

\begin{deluxetable}{ccccccccccccc}
\rotate
\tablewidth{0in}
\tabletypesize{\scriptsize}
%  \begin{center}
    \tablecaption{Best-fit parameters of spectral analysis of PCA and HEXTE
observation of GRS~1915+105 in 3-150~keV energy range during decay 2005 -- 2006 transition.
%Errors (put in parenthesis)  are given at the 99\% confidence level}
Parameter errors (put in parenthesis) correspond to 1$\sigma$ confidence level.}
    \renewcommand{\arraystretch}{1.2}
\tablehead{Observational& MJD   & $\alpha_1=\Gamma_1-1$& log($A_1$)$^\dagger$& $N_{bmc1}$, 
 & $\alpha_2=\Gamma_2-1$ &log($A_2$)$^\dagger$&   $N_{bmc2}$,  & EW,& E$_{laor}$ & Flux$^{\dagger\dagger}$& $\chi^2_{red}$ (d.o.f.)\\
ID               & day   &             &           &$L_{39}/d^2_{10}$&  &    &$L_{39}/d^2_{10}$  &  eV        &keV &  &}
 \startdata
91701-01-46-00*& 53753.43&  1.70(1)& 0.6(1)& 0.284(7) & 3.00(2) &2.0 & 0.3807(8)&108(10)&5.15$\pm$0.06& 38.60&0.93 (71)\\
91701-01-49-00*& 53771.51&  1.69(1)& 2.0& 0.221(8) & 2.41(3) &2.0 & 0.0978(2)&- &-&16.00&1.06 (71)\\
91701-01-50-00*& 53778.30&  1.65(1)& 2.0& 0.143(2) & 2.54(7)&0.38(1)& 0.1055(5)  &- &-&17.48&0.95 (71)\\
91701-01-22-00& 53794.31&  1.68(2)& 2.0& 0.143(2) & 2.00(7)&-0.79(1) & 0.081(2)  &- &-& 18.60&0.96 (73)\\
92702-01-01-00& 53803.28&  1.40(2)& 2.0& 0.12(1) & 1.8(2)   &0.29 (1)& 0.083(4)    &- &-& 19.35&1.02 (73)\\
92702-01-01-01& 53803.35&  1.50(3)& 2.0& 0.11(1) & 1.8(1)   &2.0 & 0.081(7) &- &-& 19.11&1.04 (73)\\
%92702-01-02-00*& 53809.24&  1.353(9)& 2.0& 0.128(2) & 1.74(4) &2.0 & 0.0776(4)&- &-&16.19&1.29 (73)\\
92702-01-02-01& 53809.30&  1.353(9)& 2.0& 0.085(2) & 1.74(5) &2.0 & 0.0798(4)&- &-&16.64&1.22 (73)\\
92702-01-03-00& 53815.19&  1.49(2)& 2.0& 0.094(4) & 1.21(6) &-0.20(1)& 0.063(2)     &- &-& 17.46&1.02 (73)\\
92702-01-05-00& 53829.20&  0.75(3)& 0.14(1)& 0.125(1)&   1.9(5)  &  2.0 &  0.005(3)    &- &-& 16.98&1.02 (73)\\
92082-01-05-00& 53834.84&  1.03(1)& 0.13(1)&0.140(2)&   2.0(6)  & 2.0  &  0.009(2)    &- &-& 16.05&0.98 (73)\\
92702-01-07-00& 53844.16&  1.10(2)& -0.15(2)&0.163(1)&  1.9(4)  & 2.0  &  0.005(3)    &- &-& 14.92&1.06 (73)\\
92702-01-08-00& 53851.10&  0.94(3)& 0.06(1)&0.127(2)&   2.0(3)  & 2.0 &  0.01(2)     &- &-& 14.50&1.07 (73)\\
92702-01-08-01& 53851.17&  1.13(1)& 0.09(1)&0.132(1)&   2.0(4)  & 2.0  &  0.008(3)    &- &-& 14.74&0.99 (73)\\
92702-01-08-02& 53851.23&  1.04(4)& 0.05(2)&0.128(1)&   1.9(3)  & 2.0  &  0.01(3)     &- &-& 14.03&0.95 (73)\\
90105-02-04-00& 53852.15&  1.09(2)& 0.15(1)&0.125(1)&   1.8(6)  & 2.0  &  0.009(2)    &- &-& 14.28&0.98 (73)\\
      \enddata
    \label{tab:table}
$^\dagger$
When parameter $log(A)>1.0$, this parameter is fixed to a value 2.0 (see comments in the text),     
%$^\dagger$This parameter is fixed, 
$^{\dagger\dagger}$spectral flux in the 3 -- 150 energy range in units of $\times 10^{-9}$ ergs/s/cm$^2$, 
* these data are fitted with {\it wabs*(bmc*highecut+bmc+laor+bbody)} model, see   values of the best-fit  parameters in Table 7.
%, * this observation was fitted with systematic error $>$1\%
\end{deluxetable}  

%table7

\begin{deluxetable}{cccccccccc}
\rotate
\tablewidth{0in}
\tabletypesize{\scriptsize}
%  \begin{center}
    \tablecaption{Parameters of the model: {\it wabs*(bmc*highecut+bmc+laor+bbody)} for selected  IS data (see the full list of observations in Tables 1-6).
%    Errors are given at the 99\% confidence level}
Parameter errors (put in parenthesis) correspond to 1$\sigma$ confidence level.}
    \renewcommand{\arraystretch}{1.2}
%    \begin{tabular}[h]
%      \hline
\tablehead{
Model & Parameter & 30402-01-09-00 & 90105-05-03-05 & 91701-01-04-00&91701-01-42-01  & 91701-01-43-00 &  91701-01-46-00 & 91701-01-49-00 &91701-01-50-00 }
 \startdata
bmc1 &                           &               &                 &               &              &              &             &   &    \\
     & $\alpha_1=\Gamma_1-1$     & 1.50$\pm$0.02 & 1.982$\pm$0.004 &1.981$\pm$0.005&1.91$\pm$0.01 & 1.91$\pm$0.04&1.70$\pm$0.01&1.69$\pm$0.01&1.65$\pm$0.01\\
     & kT$_1$ (keV)              & 0.9$\pm$0.1   & 1.1$\pm$0.1     &1.09$\pm$0.1   &0.99$\pm$0.05 & 1.00$\pm$0.01&1.02$\pm$0.02&0.95$\pm$0.06&0.89$\pm$0.09 \\
%     & logA$_1$                  & 7.3$^{\dagger}$&6.8$^{\dagger}$ &-0.31$\pm$0.01&              & 0.6$\pm$0.2  &             &             &    \\
     & N$_{bmc1}$                & 0.20$\pm$0.03 & 0.187$\pm$0.001 &0.185$\pm$0.002&0.129$\pm$0.003&0.154$\pm$0.015&0.284$\pm$0.007&0.221$\pm$0.008&0.143$\pm$0.002       \\
bmc2 &                           &              &               &              &              &                &               &        &   \\
     & $\alpha_2=\Gamma_2-1$     & 2.09$\pm$0.05& 2.79$\pm$0.06 &2.79$\pm$0.09 &2.70$\pm$0.06 & 2.7$\pm$0.2    &3.00$\pm$0.02  &2.41$\pm$0.03&2.54$\pm$0.07\\
     & kT$_2$ (keV)              & 1.0$\pm$0.1  & 1.0$\pm$0.1   &1.1$\pm$0.1   &1.0$\pm$0.2   & 0.99$\pm$0.09  &1.0$\pm$0.2    &0.99$\pm$0.08&1.1$\pm$0.3  \\
     & N$_{bmc2}$                & 0.11$\pm$0.01&0.112$\pm$0.001&0.110$\pm$0.001&0.1644$\pm$0.0005&0.14$\pm$0.09&0.3807$\pm$0.008&0.0978$\pm$0.0002&0.1055$\pm$0.0005\\
laor &                           &             &                &               &              &               &               &           &  \\
     & E$_{laor}$ (keV)          & 5.7$\pm$0.3& 6.46$\pm$0.01   & 6.33$\pm$0.02 &5.39$\pm$0.01 & 5.39$\pm$0.01 &5.15$\pm$0.06  &6.40$\pm$0.09&6.40$\pm$0.02 \\
     & N$_{laor}$                & 0.073$\pm$0.02 & 0.062$\pm$0.001&0.028$\pm$0.002&0.06$\pm$0.01&0.06$\pm$0.01&0.022$\pm$0.002&0.016$\pm$0.004&0.016$\pm$0.009 \\
     & EW$_{laor}$ (eV)          & 290$\pm$18     &  378$\pm$10 & 256$\pm$15   &150$\pm$15     & 180$\pm$10    &108$\pm$10     &100$\pm$10 & 100$\pm$10   \\
     &  (eV)        & & &   & &    &   & &    \\
``bbody'' &                          &             &                &             &            &               &                &   &  \\
     & T$_{"bbody"}$          & 4.1$\pm$0.2 & 4.5$\pm$0.3    & 4.3$\pm$0.3 &4.4$\pm$0.2 & 4.68$\pm$0.09 &4.77$\pm$0.04   &4.47$\pm$0.09&4.5$\pm$0.1 \\
     &  (keV)        & & &   & &    &   & &    \\
     & N$_{"bbody"}$               & 0.04$\pm$0.01& 0.011$\pm$0.006&0.036$\pm$0.006&0.041$\pm$0.005&0.009$\pm$0.001&0.024$\pm$0.006&0.038$\pm$0.003&0.048$\pm$0.004   \\
     & EW$_{"bbody"}$         &550$\pm$30&430$\pm$20 & 490$\pm$40   &710$\pm$50& 390$\pm$30    &470$\pm$30    &540$\pm$30 & 630$\pm$40   \\
     &  (eV)        & & &   & &    &   & &    \\
Flux$^{\dagger}$ &               &             &           &                  &            &          &            & &    \\
     &                           & 23.89       & 22.67     & 22.69            &20.57       &20.29     &  38.60     &  16.00 & 17.48   \\
%      \hline
     & $\chi^2$ (d.o.f.)         & 1.04  (71)  & 0.96 (71) & 0.97 (71)        &1.01 (71)   & 0.92 (71)&  0.93 (71) & 1.06 (71)  &  0.95 (71) \\
S$_{15GHz}$ $^{\dagger\dagger}$&                    & 88$\pm$2    & 156$\pm$2 & 140$\pm$2        &41$\pm$2    &51$\pm$2  &65$\pm$2&40$\pm$2 & 20$\pm$2      \\
      \hline
      \enddata
%      \hline
%      \end{tabular}
    \label{tab:par_bbody}
%  \end{center}
%Parameters not listed in the table were fixed
%$^\dagger$this parameter is fixed, 
$^{\dagger}$spectral flux in the 3
-- 150 energy range in units of $\times 10^{-9}$ erg/s/cm$^2$.
$^{\dagger\dagger}$spectral flux density in radio band centered at 15 GHz in units of mJy.
\end{deluxetable}

\clearpage
\newpage
\begin{figure}[ptbptbptb]
\includegraphics[scale=1.,angle=0]{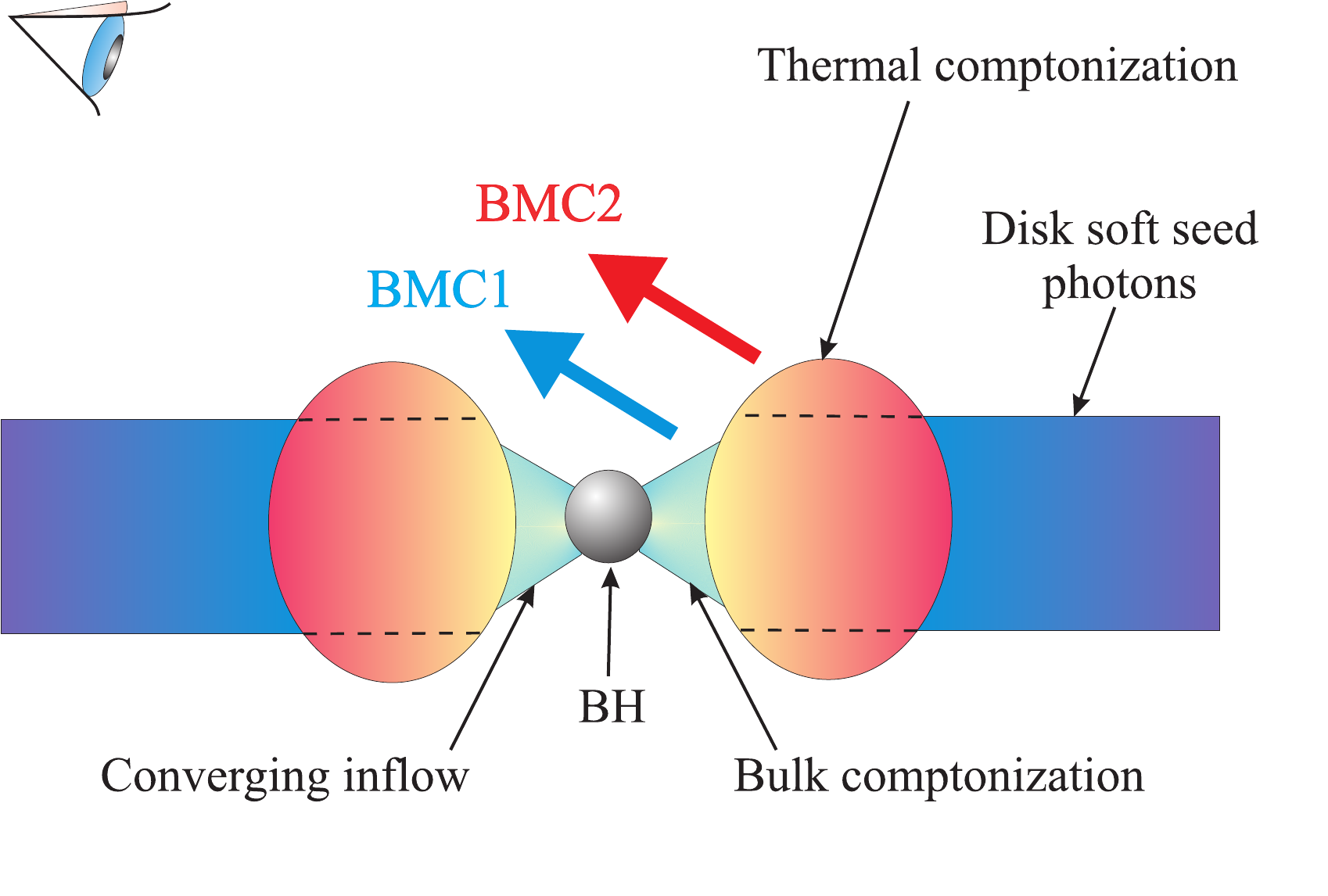}
\caption{ Schematic view of the proposed geometry for thermal and bulk Comptonization regions in the source hosting a BH with PL-like emission at high energies.  The thermal
plus bulk Comptonization spectrum (thermal plus
bulk {\it BMC1}) arises in the innermost part of the transition layer (TL), where the disk BB-like  seed photons are (thermally and
dynamically) Comptonized by the in-falling material. Whereas the
thermal Comptonization spectrum (thermal {\it BMC2}) originates in the outer part of the TL region.
%The
%radial extents of the two Comptonization regions are, in fact, rescaled for clarity, whereas numerical calculations show that they are of the order of $3R_{\rm S}$.
}
\label{geometry}
\end{figure}

\begin{figure}[ptbptbptb]
\includegraphics[scale=0.9,angle=0]{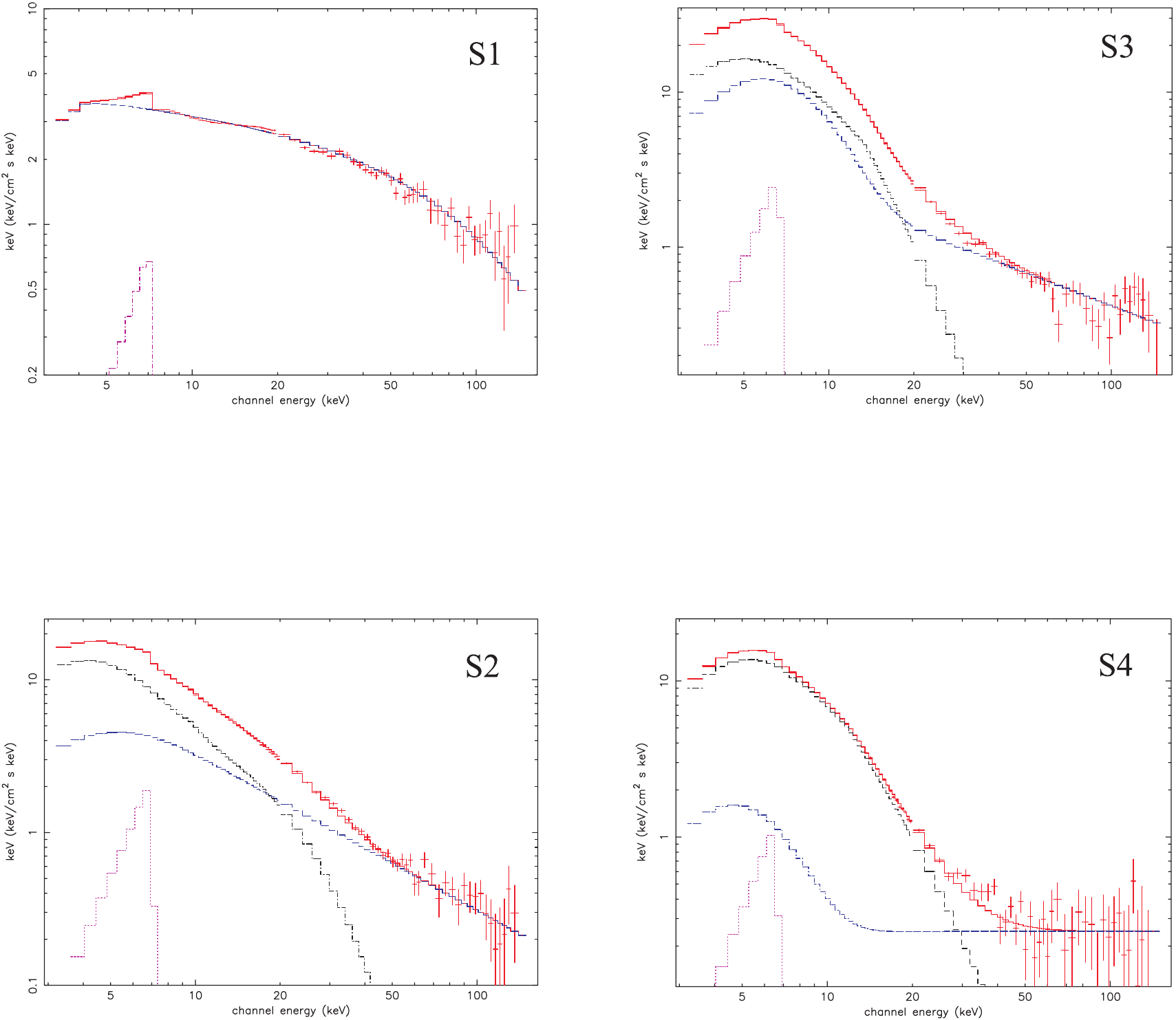}
\caption{Evolution of spectum shape of  GRS~1915+105 during
LHS, IS, HSS  and VSS spectral states. 
Data are taken from \textit{RXTE} observations 
20402-01-11-00 ({\it S1}, $\Gamma_1=1.8$, LHS), 
91701-01-33-00 ({\it S2}, $\Gamma_1=2.9$, $\Gamma_2=3.7$, IS), 
91701-01-11-00 ({\it S3}, $\Gamma_1=2.7$, $\Gamma_2=4.1$, HSS) and 
91701-01-19-00 ({\it S4}, $\Gamma_2=4.2$, VSS). 
Here data are denoted by red points,
the spectral  model presented with components are shown by 
 blue, black    and 
 dashed purple  lines for {\it BMC1}, {\it BMC2} and {\it laor} components  respectively.
}
\label{spec_evol_4Sm}
\end{figure}

%\begin{figure}[ptbptbptb]
%\includegraphics[scale=1.0,angle=0]{del_1.eps}
%\includegraphics[scale=0.65,angle=-90]{f2b.eps}
%\caption{Fit of IS-HSS spectrum (observation 91701-01-11-00, on 18 May 2005) with  models which include  a single BMC component.  
%Left panel: $wabs*bmc*highecut$  ($\chi^2_{red}$=12.3 for 75 d.o.f.)  and right panel:$wabs*(bmc*highecut+laor)$ ($\chi^2_{red}$=12.1 for 77 d.o.f.).
%The count spectrum is presented in units of  counts/s$^{-1}$cm$^{-2}$.
%Bottom panels   show $\Delta\chi$. 
%}
%\label{one_BMC}
%\end{figure}

\begin{figure}[ptbptbptb]
\includegraphics[scale=0.8,angle=0]{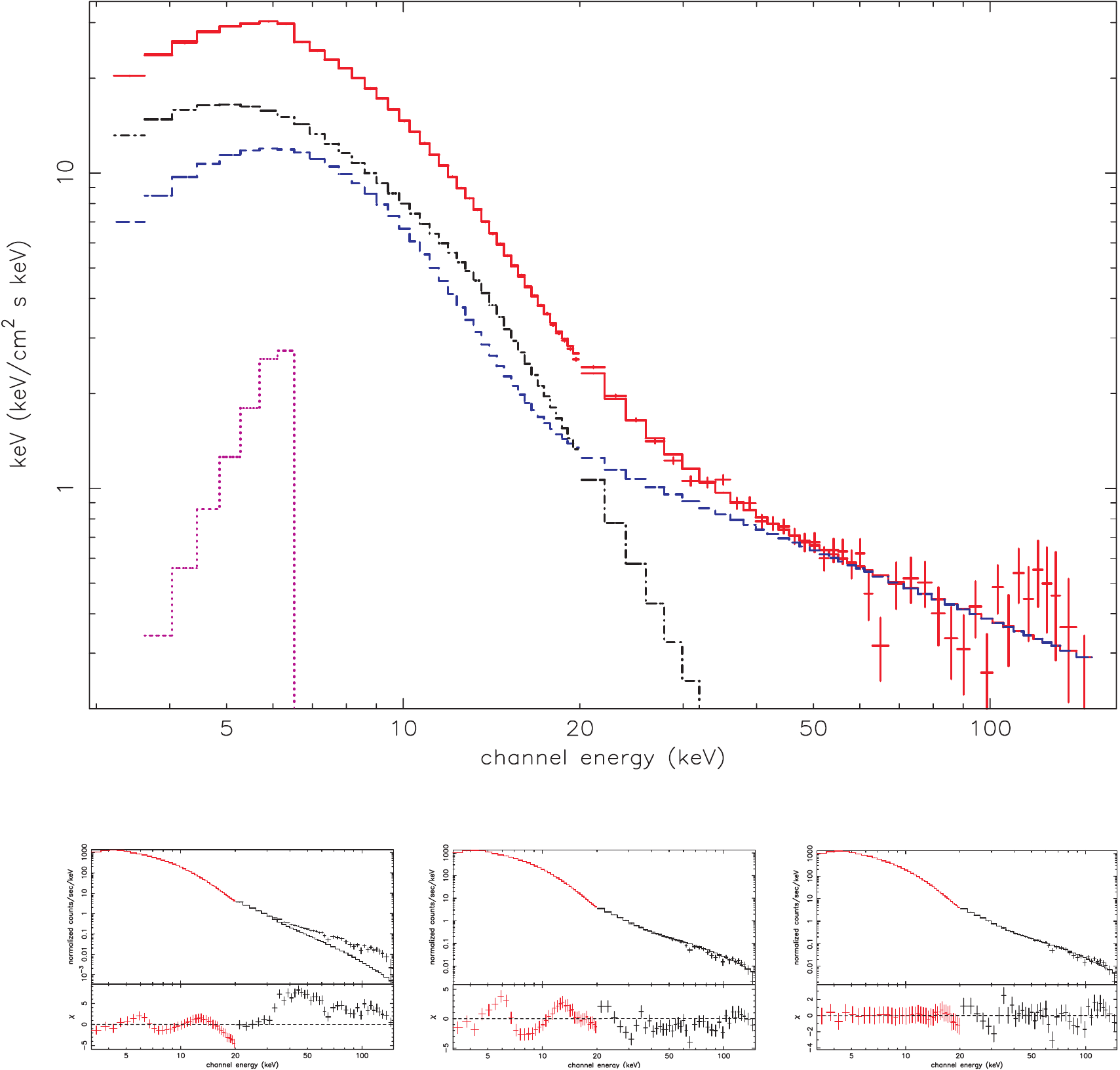}
\caption{The best-fit spectrum of GRS~1915+105 during HSS events of 2005 
%outburst 
rise transition 
(observation 91701-01-11-00 ) in  $E*F(E)$ units ({\it top}) along with the count spectrum and 
$\Delta\chi$ ({\it bottom panels}). 
{\it Bottom panels: Left}: the best-fit spectrum and $\Delta\chi$ for the model  fit,  $wabs*bmc*highecut$, which  include a single BMC component without an  iron line  component ($\chi^2_{red}$=12.3 for 75 d.o.f.), 
{\it center:} the model fit, $wabs*(bmc+bmc*highecut)$,  which includes  two BMC components 
 without an iron line component ($\chi^2_{red}$=3.28 for 
 76 d.o.f.) and {\it right:} the same as  the latter  one but adding of  the {\it Laor} line component, 
 $wabs*(bmc+bmc*highecut+laor)$,
 ($\chi^2_{red}$=1.04 for 73 d.o.f., see Table 4).
% Fit of IS-HSS spectrum  with  models which include  two BMC components.  
%Top: The spectrum and model  presented with its components in $E*F(E)$ units.  
%Bottom panels   show  count spectrum and $\Delta\chi$. 
%On the top 
 %panel data are denoted by red points, the model is shown along with components.  
 Blue, black, purple  lines stand  for {\it BMC1},  {\it BMC2}  and  {\it Laor} components respectively.
%Left panel: $wabs*(bmc+bmc*highecut)$  ($\chi^2_{red}$=3.28 for 
 %76 d.o.f.) and right panel:$wabs*(bmc+bmc*highecut+laor)$ ($\chi^2_{red}$=1.04 for 73 d.o.f.,  
 %see Table 4).
}
\label{two_BMC}
\end{figure}

%\begin{figure}[ptbptbptb]
%\includegraphics[scale=1.,angle=0]{ff2.eps}
%\includegraphics[scale=0.65,angle=-90]{f2b.eps}
%\caption{Three representative $EF_{E}$ spectral diagrams  during the rise part of the 
%1997 outburst of GRS~1915+105. Data are taken from \textit{RXTE} observations 
%20402-01-19-00 ({\it blue}, $\Gamma_1=1.6$), 
%20402-01-30-01 ({\it green}, $\Gamma_1=2.1$) and 
%20402-01-53-00 ({\it red}, $\Gamma_1=3.0$). 
%}
%\label{spec_evol_97_rise}
%\end{figure}

\begin{figure}[ptbptbptb]
\includegraphics[scale=0.9,angle=0]{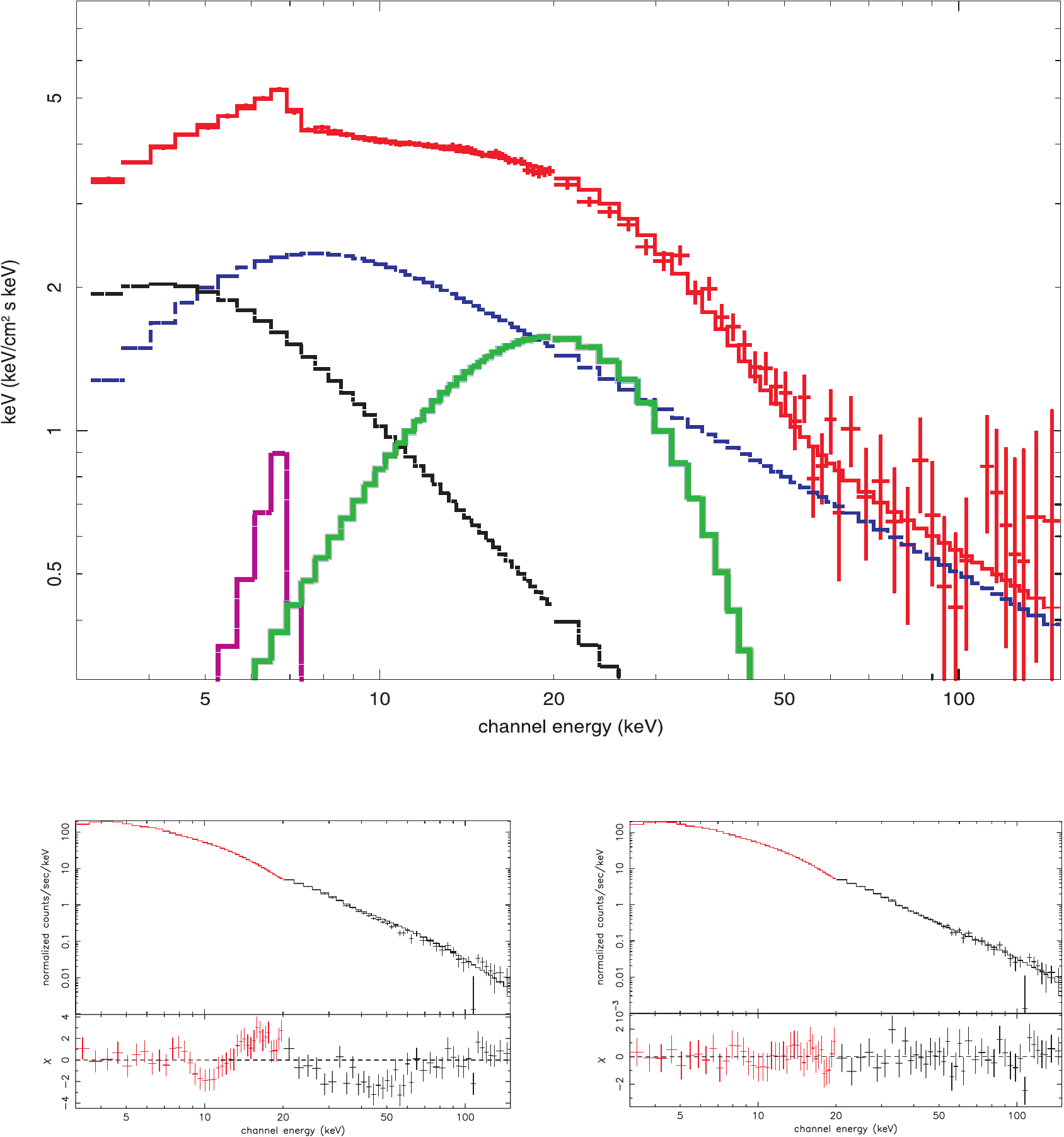}
\caption{
The best-fit spectrum during IS-LHS of 2005-2006 
%outburst 
decay transition 
in $EF(E)$ units ({\it top}) and in normalized 
counts units ({\it bottom panel}) with $\Delta\chi$ for the 91701-49-00 
observation. {\it Left}: fitting without modelling the high-temperature
``bbody''  component
($\chi^2_{red}$=1.43 for 73 d.o.f.) and {\it right}: the best-fit spectrum and $\Delta\chi$, when 
the bump in 
residuals  at $\sim 20$ keV is modelled by a ``high-temperature  bbody-like''
component with  $\chi^2_{red}$=1.06 for 70 d.o.f.. On top
panel the data are denoted by red points, 
the spectral  model presented with components are shown by 
 blue, black, 
  purple  and  green lines for {\it BMC1}, {\it BMC2},  {\it laor} and ``high-temperature'' {\it bbody} components  respectively.
}
\label{sp_bbody}
\end{figure}

\begin{figure}[ptbptbptb]
\includegraphics[scale=1.,angle=0]{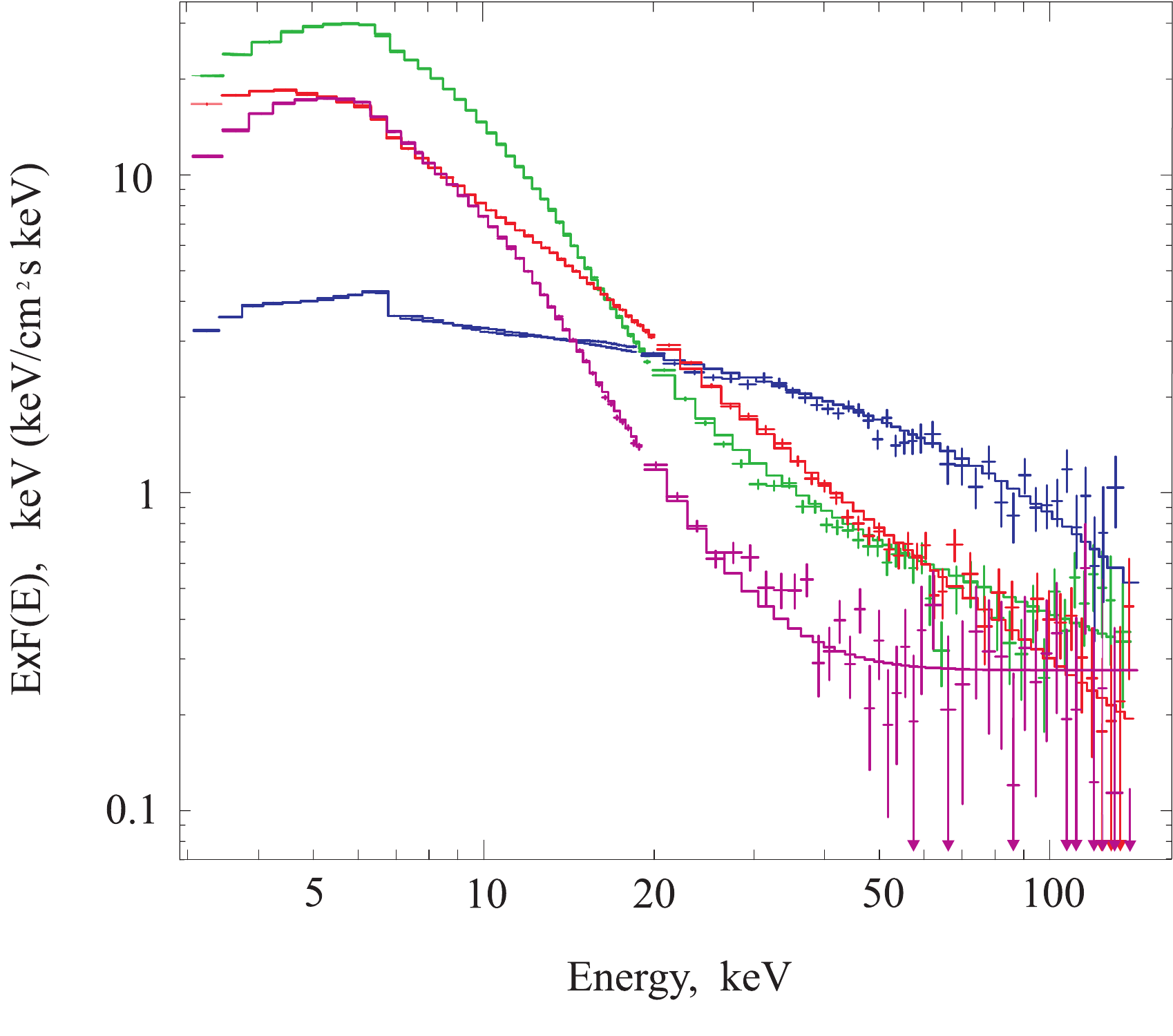}
\caption{Four representative $EF_{E}$ spectral diagrams  during 
LHS, IS, HSS and VSS spectral states of GRS~1915+105. 
Data are taken from \textit{RXTE} observations 
20402-01-11-00 ({\it blue},  LHS), 
91701-01-33-00 ({\it red},  IS), 
91701-01-11-00 ({\it green}, HSS) and 
91701-01-19-00 ({\it purple},  VSS). 
}
\label{spec_evol_4S}
\end{figure}

\begin{figure}[ptbptbptb]
\includegraphics[scale=0.65,angle=0]{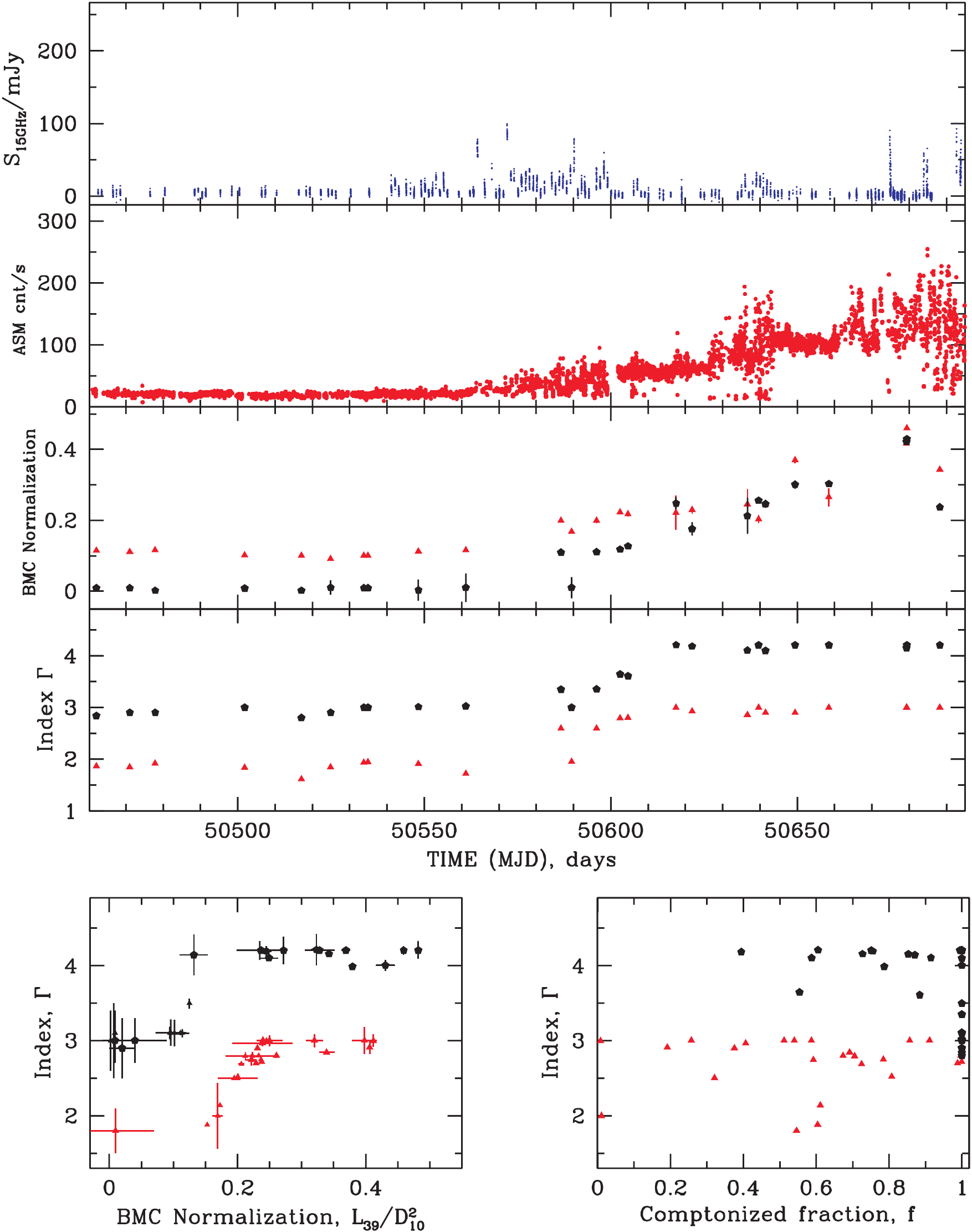}
\caption{{\it From Top to Bottom:}  
Evolutions of the flux density $S_{15GHz}$ at 15 GHz (Ryle Telescope), 
\textit{RXTE}/ASM count rate, 
BMC normalization and 
photon index $\Gamma$ 
in the beginning of the 1997 
%outburst 
rise transition of GRS~1915+105 (MJD 50460-50700).
Red/black points ({\it for  two last panels}) correspond 
to hard/soft components with $\Gamma_1$ and $\Gamma_2$, respectively. 
{\it Bottom:} Spectral index $\Gamma$ plotted versus BMC normalization ({\it left}) 
and Comptonized fraction ({\it right}) for this  transition. Here the red 
triangles/black circles correspond to hard/soft components with 
$\Gamma_1$ and $\Gamma_2$, correspondingly. Note that in most cases the normalization of the hard BMC component ($BMC1$) is higher than that of the soft component ($BMC2$) (see red points vs black points in BMC normalization-time panel). 
}
\label{outburst_97_rise}
\end{figure}

\begin{figure}[ptbptbptb]
\includegraphics[scale=0.65,angle=0]{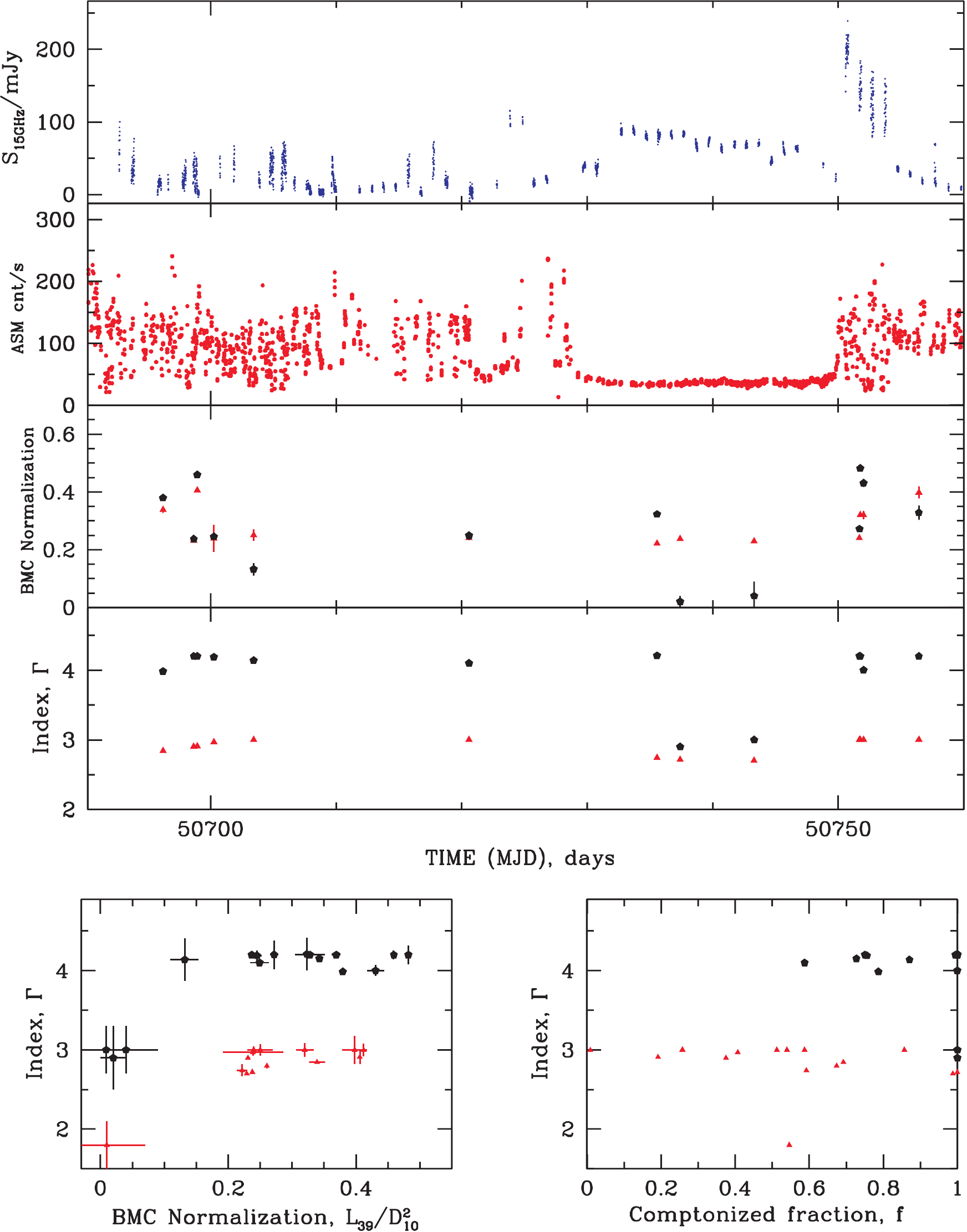}
\caption{{\it From Top to Bottom:} 
 The  same types of evolutions which are   presented in Fig.  \ref{outburst_97_rise}
 %of the flux density $S_{15GHz}$ at 15 GHz (Ryle Telescope), 
%ASM/RXTE count rate, 
%BMC normalization and 
%photon index $\Gamma$ 
but now that  for the  middle  of the 1997 
%outburst 
rise transition of GRS~1915+105 (MJD 50650-50760). 
%Red triangles/black circles ({\it for the two last panels}) correspond  
%to hard/soft components with $\Gamma_1$ and $\Gamma_2$, respectively. 
%{\it Bottom:} Spectral index plotted versus BMC normalization ({\it left}) 
%and Comptonized fraction ({\it right}).
%Here red triangles/black circles marked hard/soft components, correspondingly.
In most cases the normalization of the soft BMC component ($BMC2$) is higher than that of the hard component ($BMC1$) [see black  points vs red points in BMC normalization - time panel and compare with that in 
Fig.  \ref{outburst_97_rise}].  The pivoting effect occurs (i.e. 
$N_{bmc1} >N_{bmc2}$ 
switches to  $N_{bmc2} >N_{bmc1}$  around  MJD 50690. 
}
\label{outburst_97_middle}
\end{figure}

\begin{figure}[ptbptbptb]
\includegraphics[scale=0.65,angle=0]{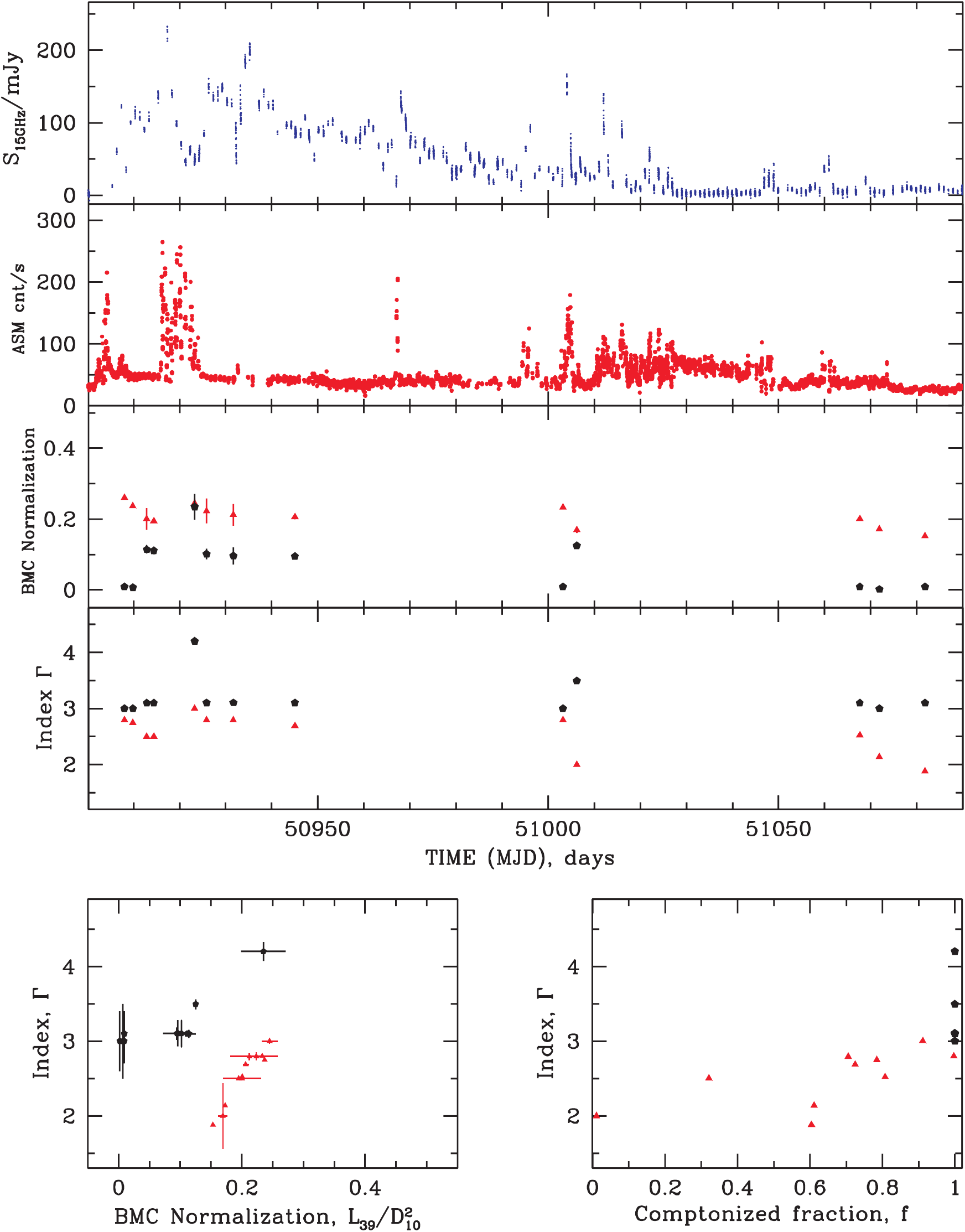}
\caption{{\it From Top to Bottom:}  
The  same types of evolutions which are   presented in Figs.~\ref{outburst_97_rise}-\ref{outburst_97_middle}
%Evolution of the flux density $S_{15GHz}$ at 15 GHz (Ryle Telescope), 
%ASM/RXTE count rate, 
%BMC normalization and 
%photon index $\Gamma$ 
but now that for  the 1997 -- 1998 decay transition of GRS~1915+105.
In most cases the normalization of the hard BMC component ($BMC1$) is higher than that of the soft component ($BMC2$) (see red points vs black points in BMC normalization-time panel). 
The pivoting effect occurs  around  MJD 50910. 
%Red/black points ({\it for the two last panels}) correspond 
%to hard/soft components with $\Gamma_1$ and $\Gamma_2$, respectively. 
%{\it Bottom:} Spectral index $\Gamma$ plotted versus BMC normalization ({\it left}) 
%and Comptonized fraction ({\it right}) for this  transition. Here red/black points correspond to hard/soft components with 
%$\Gamma_1$ and $\Gamma_2$, correspondingly.
}
\label{outburst_97-98_decay}
\end{figure}

\begin{figure}[ptbptbptb]
\includegraphics[scale=0.65,angle=0]{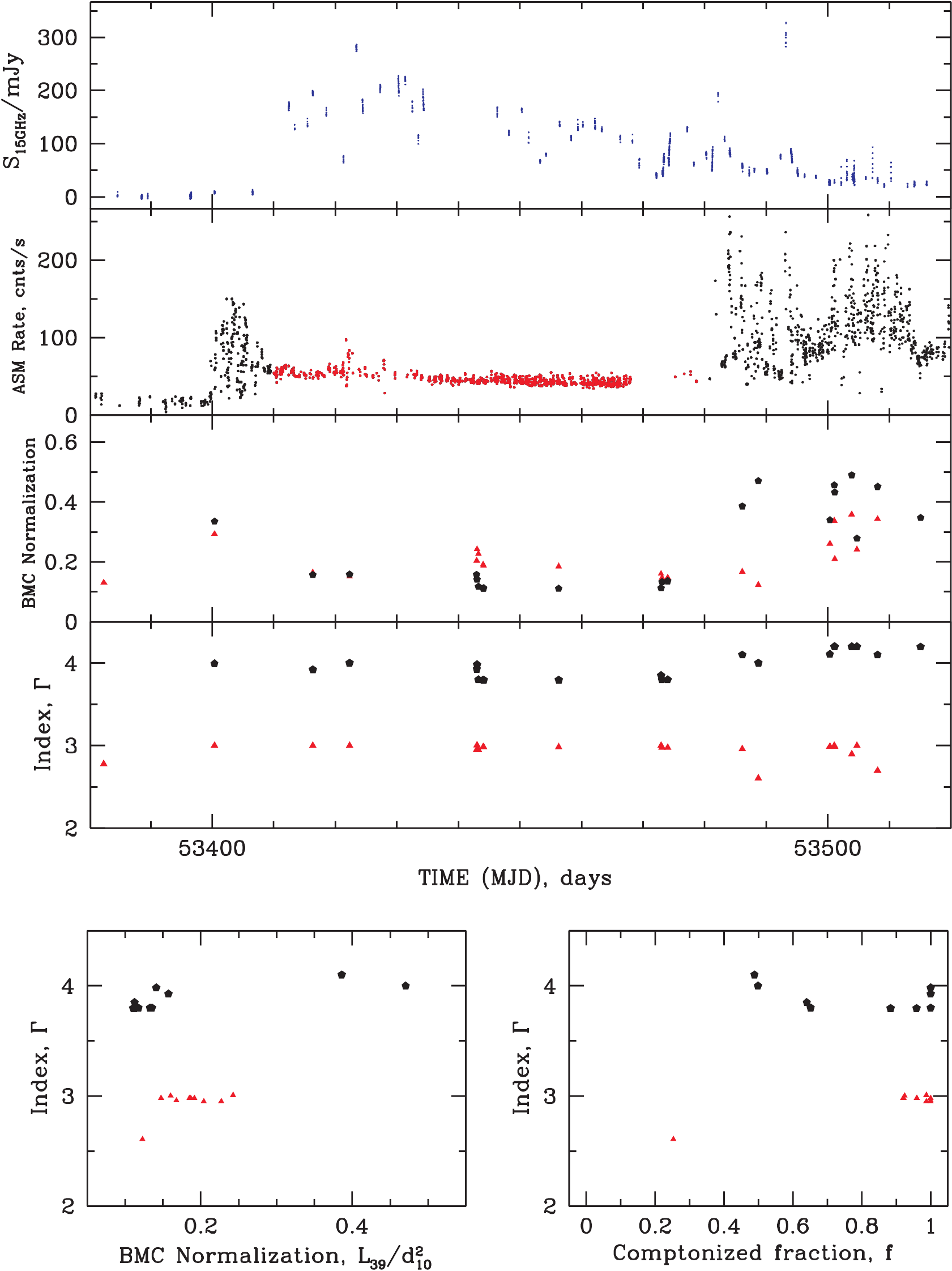}
\caption{{\it From Top to Bottom:}  
The  same types of evolutions which are   presented in Figs.  \ref{outburst_97_rise}-\ref{outburst_97-98_decay}
but now that  for  the  2005  rise  transition of GRS~1915+105.
The pivoting effect occurs  around  MJD 53475. 
%Evolution of the flux density $S_{15GHz}$ at 15 GHz (Ryle Telescope), 
%ASM/RXTE count rate,  BMC normalization and photon index $\Gamma$ 
%during the  2005 outburst  transition of GRS~1915+105.
%Red points on the {\it second} panel mark  {\it Intermediate state}.
%Red/black points ({\it for the two  last panels}) correspond  
%to hard/soft BMC spectral components with $\Gamma_1$ and $\Gamma_2$, respectively. 
%{\it Bottom:} Spectral index plotted versus BMC normalization ({\it left}) 
%and Comptonized fraction ({\it right}).
% for the {\it Intermediate state}. 
%Here black/red points mark hard/soft BMC components, correspondingly. 
}
\label{outburst_05_IS}
\end{figure}

%\begin{figure}[ptbptbptb]
%\includegraphics[scale=0.65,angle=0]{f4.eps}
%\caption{ {\it From Top to Bottom:}  
%Evolution of flux density $S_{15GHz}$ at 15 GHz (Ryle Telescope), 
%ASM/RXTE count rate, 
%BMC normalization and 
%photon index $\Gamma$ 
%during 2005 outburst  which includes the decay transition.
%Red/black circles ({\it for two last panels}) correspond 
%to soft/hard spectral components with $\Gamma_1$/$\Gamma_2$, respectively. 
%{\it Bottom:} Spectral index $\Gamma$ plotted versus BMC normalization ({\it left}) 
%and Comptonized fraction ({\it right}).  Here red 
%/black point correspond to hard/soft spectral components with 
%$\Gamma_1$/$\Gamma_2$, respectively.
%}
%\label{outburst_05_decay}
%\end{figure}

\begin{figure}[ptbptbptb]
\includegraphics[scale=0.65,angle=0]{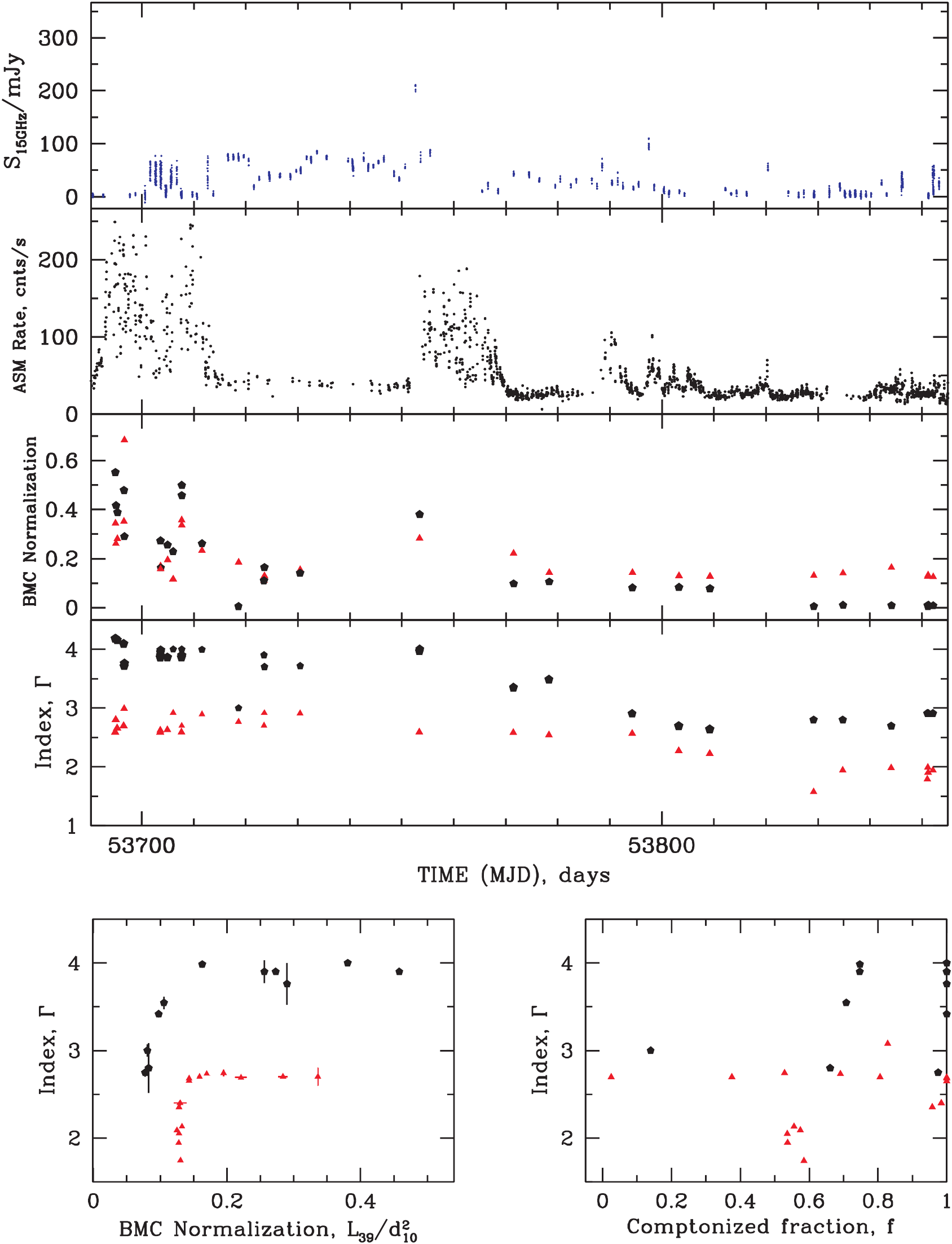}
\caption{ {\it From Top to Bottom:}   
The  same types of evolutions which are   presented in Figs.~\ref{outburst_97_rise}-\ref{outburst_05_IS}
but now that  for  the 2005 -- 2006  middle and decay transitions.  
The pivoting effect occurs  between   MJD 53750 and  MJD 53775.
%Evolution of the flux density $S_{15GHz}$ at 15 GHz (Ryle Telescope), 
%ASM/RXTE count rate, 
%BMC normalization and 
%photon index $\Gamma$ during the 2005-2006  outburst middle and decay transition. 
%Red/black circles ({\it for the two last panels}) correspond 
%to soft/hard spectral components with $\Gamma_1$ and $\Gamma_2$, respectively. 
%{\it Bottom:} Spectral index $\Gamma$ plotted versus BMC normalization ({\it left}) 
%and Comptonized fraction ({\it right}).  Here red 
%/black point correspond to hard/soft spectral components with 
%$\Gamma_1$ and $\Gamma_2$, respectively.
}
\label{outburst_05-06_decay}
\end{figure}

\begin{figure}[ptbptbptb]
\includegraphics[scale=0.9,angle=0]{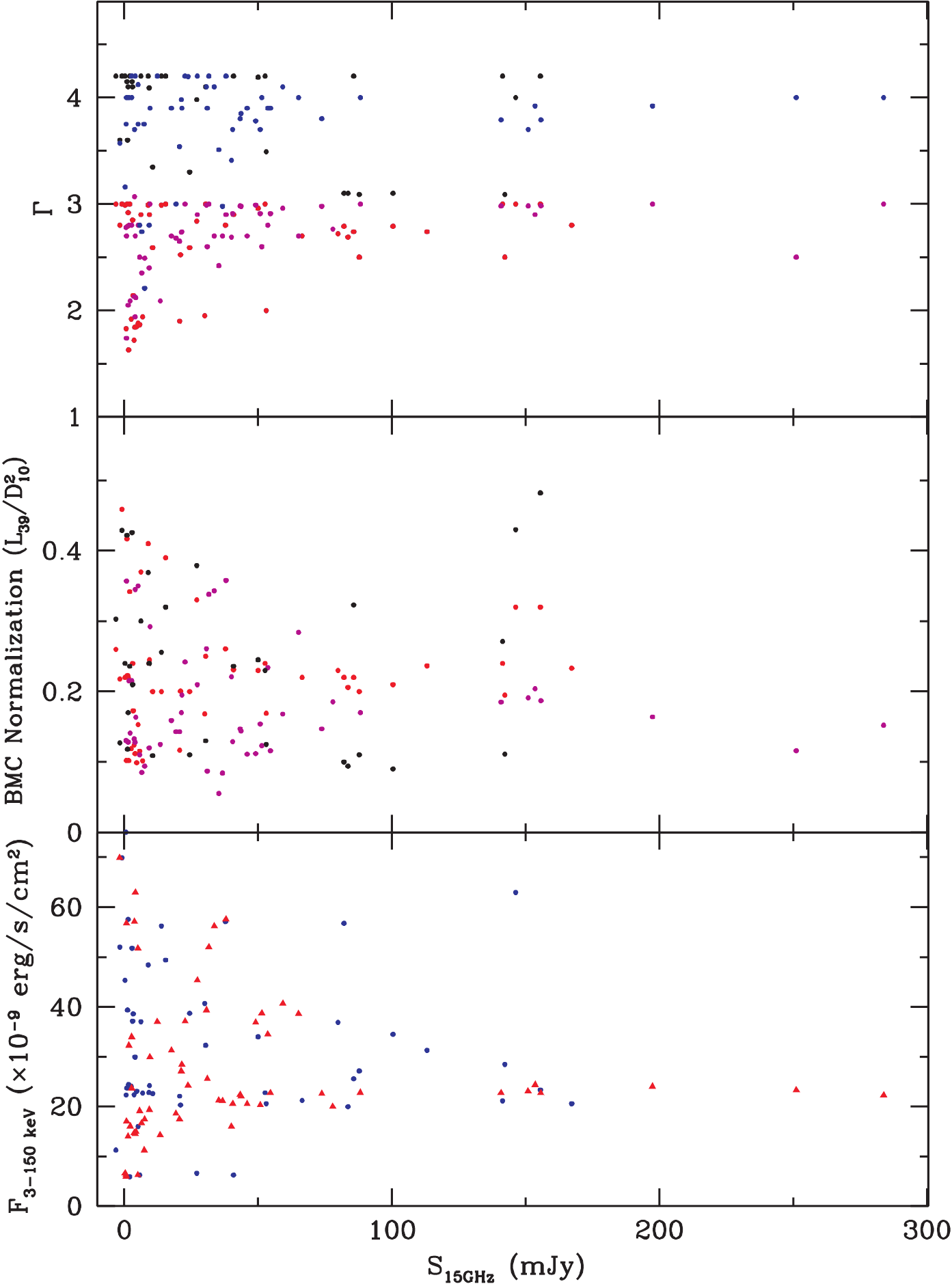}
\caption{ BMC photon  indices $\Gamma$ ({\it top}), normalization ({\it middle}) and X-ray 
flux ({\it bottom}) vs flux density $S_{15GHz}$. {\it Top and middle:} Red and black points 
stand for hard and soft spectral components respectively for the 1997 -- 1998 active episode. Crimson 
and blue points correspond to hard and soft spectral components respectively for 
the 2005 -- 2006 episode. {\it Bottom:} Blue circles and  red triangles stand for 1997 -- 1998 and 2005 -- 2006.
%while  correspond to  outburst. 
}
\label{outburst_index_radio_X-ray}
\end{figure}

\begin{figure}[ptbptbptb]
\includegraphics[scale=0.9,angle=0]{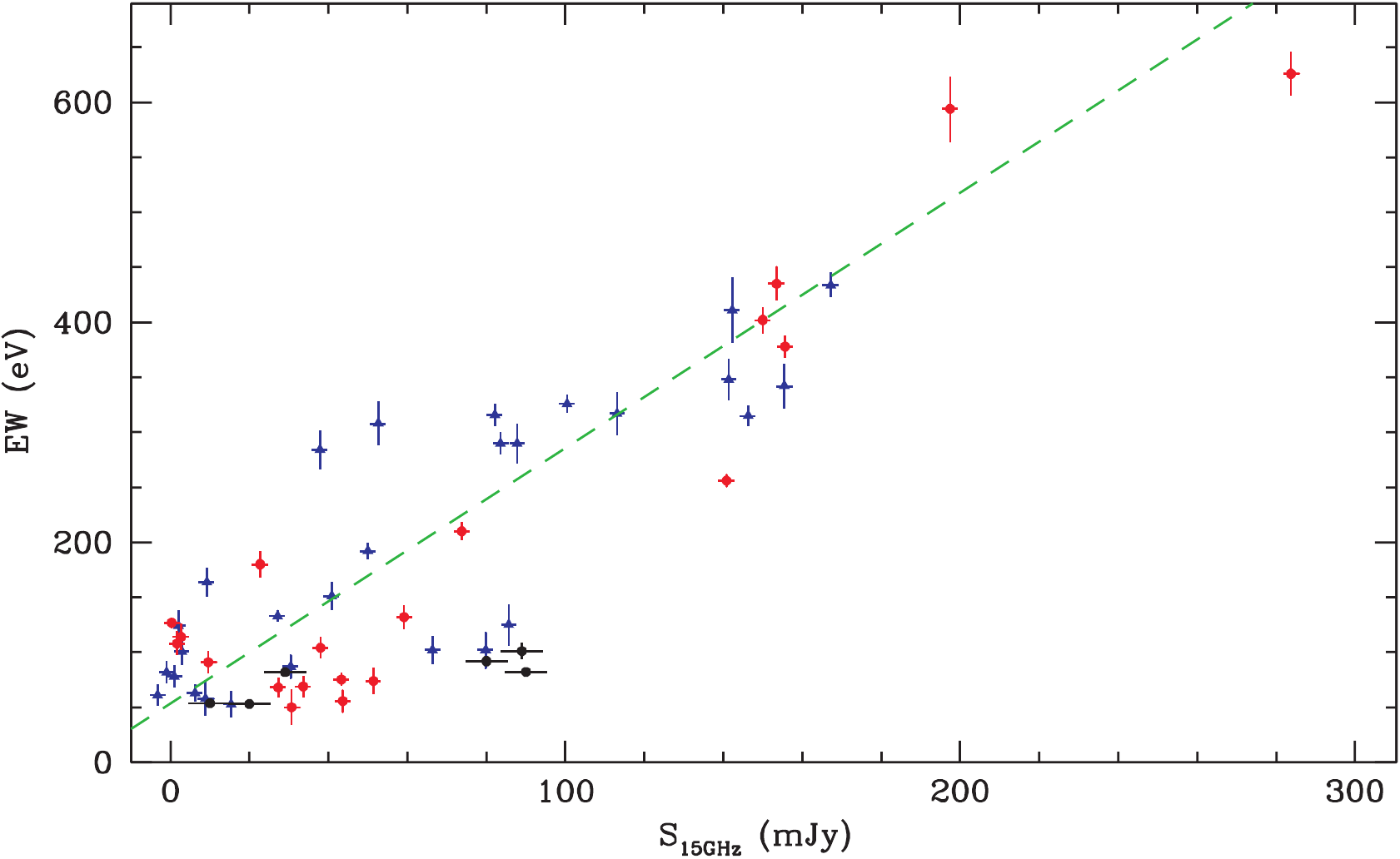}
\caption{The equivalent width (EW) of iron line in eV as a function of flux density 
$S_{15GHz}$ at 15 GHz (Ryle Telescope) in mJy for 1997/2005 data sets (blue/red
points).  Here we also include  black points   which have been recently found by 
\cite{nl09} analyzing archival HETGS (High Energy Transmission Grating Spectrometer) observations of
GRS 1915+105 from the Chandra X-ray Observatory. }
\label{outburst_EW_radio}
\end{figure}
%\begin{figure}[ptbptbptb]
%\includegraphics[scale=0.65,angle=0]{f5.eps}
%\includegraphics[scale=0.65,angle=-90]{f2b.eps}
%\caption{Evolution of flux density $S_{15GHz}$ at 15 GHz (Ryle Telescope), 
%ASM/RXTE count rate,  BMC normalization and photon index $\Gamma$ 
%during 2005 outburst rise transition of GRS~1915+105.
%Red points on the {\it second} panel mark  {\it Intermediate state}.
%Red triangles/black circles ({\it for two last panels}) correspond  
%to hard/soft BMC spectral components with $\Gamma_1$/$\Gamma_2$, respectively. 
%{\it Bottom:} Spectral index plotted versus BMC normalization ({\it left}) 
%and Comptonized fraction ({\it right}) for the {\it Intermediate state}. 
%Here triangles/circles mark hard/soft BMC components, correspondingly. 
%}
%\label{outburst_05_IS}
%\end{figure}

\begin{figure}[ptbptbptb]
\includegraphics[scale=0.9,angle=0]{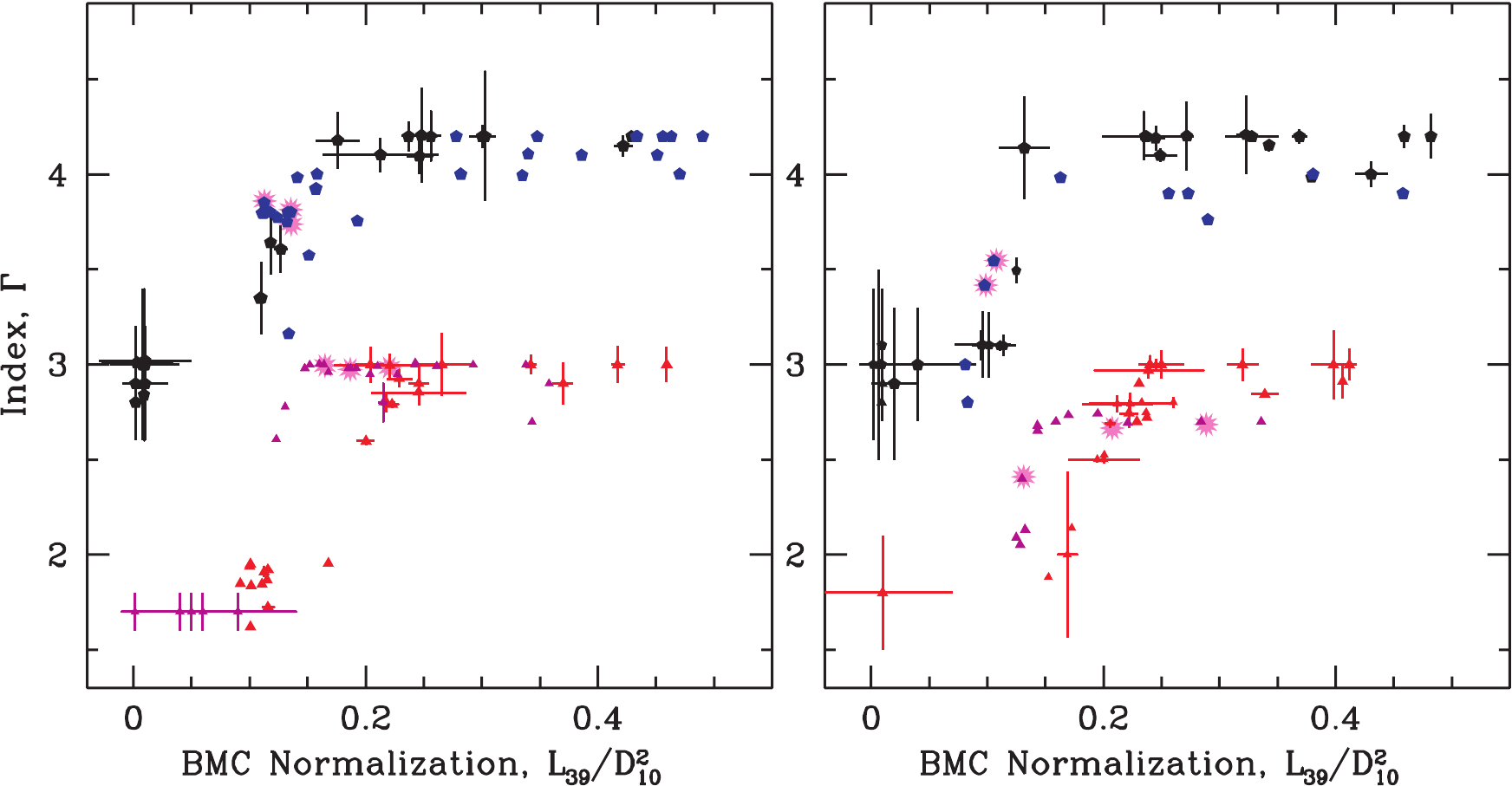}
\caption{Photon index plotted versus BMC normalization for rise 
({\it left}) and decay ({\it right})  transitions.
Red/black points  stand for  hard/soft spectral components 
 for the 1997 activity episode.
 % compared with those for  1997 outburst rise.
Crimson  and blue points correspond to hard and soft spectral components
respectively for the 2005 episode. Points marked with rose oreol correspond to IS-HSS spectra fitted by
the model which includes  ``high-temperature bbody'' component (for details see Fig. \ref{sp_bbody} and Table 7).
}
\label{outburst_index_norm}
\end{figure}

\begin{figure}[ptbptbptb]
\includegraphics[scale=0.9,angle=0]{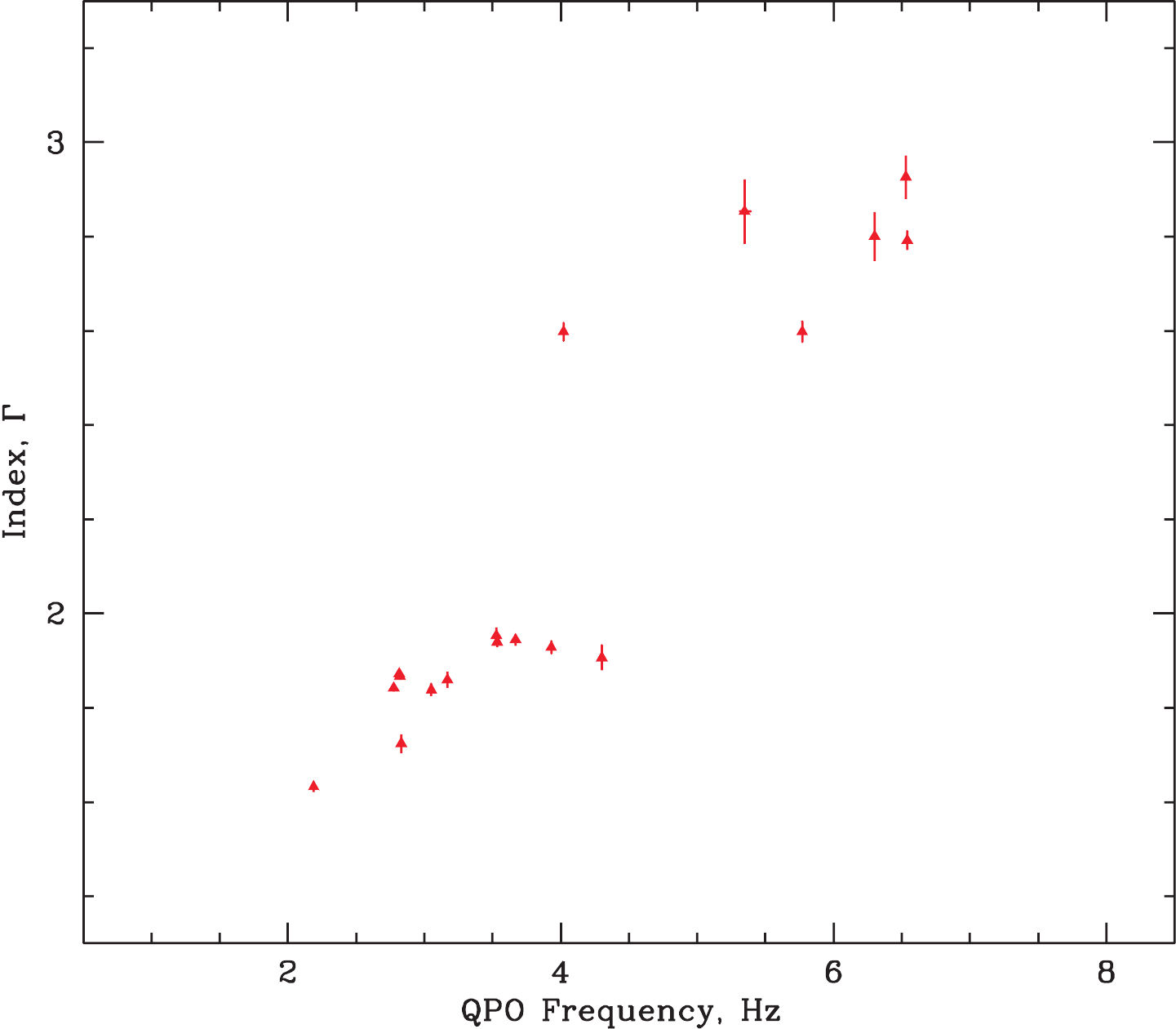}
\caption{Photon index $\Gamma_1$ (hard component) plotted versus QPO 
centroid for the 1997 rise transition from GRS~1915+105. 
%Green circles mark sub-harmonic points. Blue circles mark cases with two 
%QPO in power spectrum of the object.
}
\label{outburst_index_qpo}
\end{figure}

\begin{figure}[ptbptbptb]
\includegraphics[scale=0.8,angle=0]{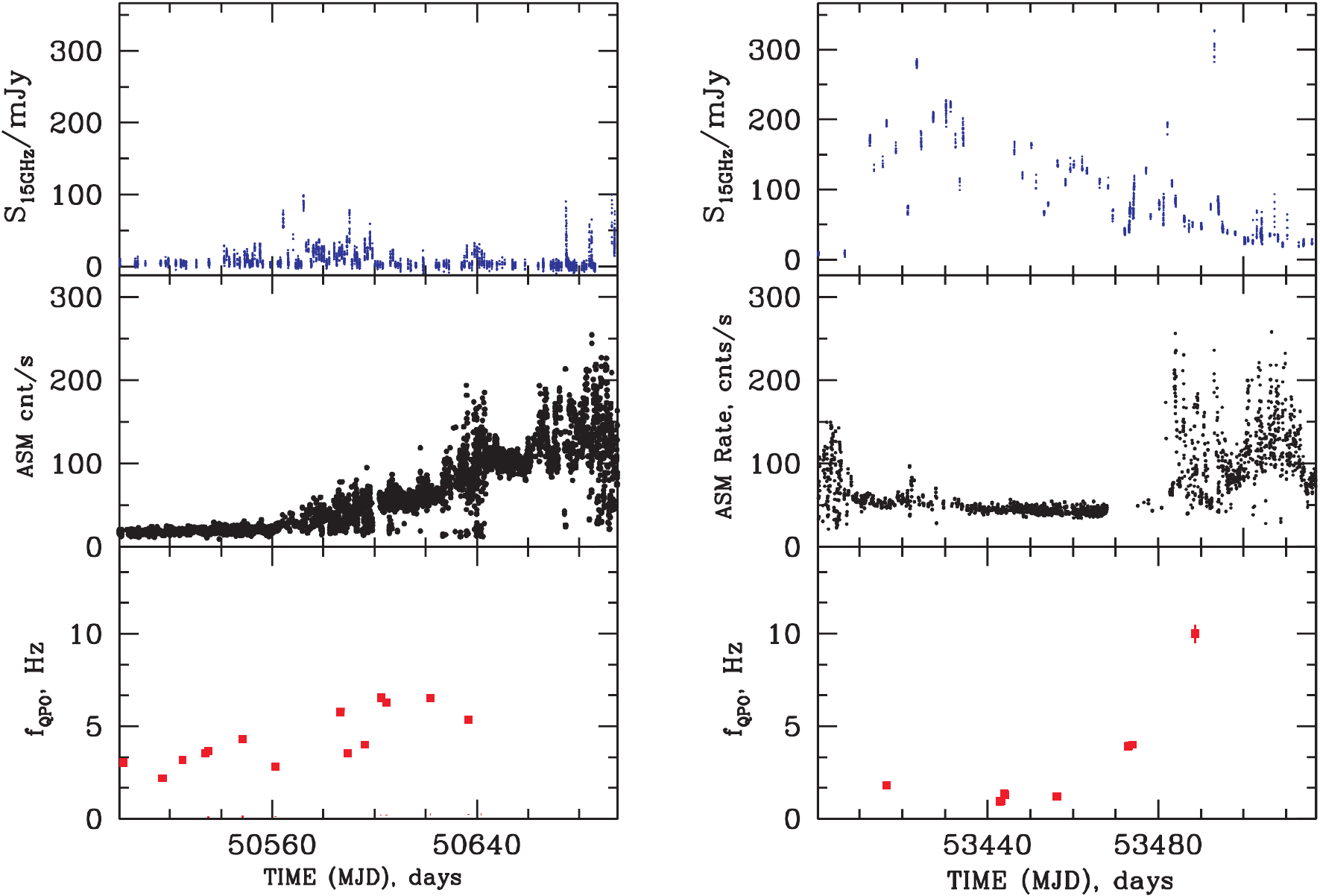}
\caption{
Evolution of flux density $S_{15GHz}$ at 15 GHz (Ryle Telescope), 
\textit{RXTE}/ASM count rate, 
$\nu_{QPO}$ 
during 1997 (left column) and 2005 (right column) rise transitions of GRS~1915+105. 
Here $\nu_{QPO}$ stands for  the centroid frequency of the fundamental QPO. 
The left column panel demonstrates the presence of QPO even when for radio  flux is low  ($<$30
mJy). 
The right column panel shows an example of the presence of QPOs during prominent radio flux events 
($\sim$100 -- 200 mJy). These panels demonstrate that  the QPO
appearances are independent of the radio flux during LHS-HSS transitions.
% from radio events of GRS~1915+105. 
}
\label{radio_QPO_independence}
\end{figure}

%\begin{figure}[ptbptbptb]
%\includegraphics[scale=0.6,angle=-90]{1915vs1550norm.eps}
%\includegraphics[scale=0.9,angle=0]{f5_old.eps}
%\includegraphics[width=5in,height=5.in,angle=0]{f5.eps}
%\caption{Scaling of photon index vs BMC normalization correlation for   GRS~1915+105 (red) with that 
%for  XTE 1550-564 (black). Scaling  ratio $C_{N,1550}(\Gamma)/C_{N,1915}(\Gamma)=3.66\pm 0.29$.
%Curves are the best-fit of the index-normalization correlations}
%\label{scaling_index_norm}
%\end{figure}

%\begin{figure}[ptbptbptb]
%\includegraphics[scale=0.8,angle=0]{f7.eps}
%\includegraphics[width=5in,height=5.in,angle=0]{f5.eps}
%\caption{Evolution of flux density $S_{15GHz}$ at 15 GHz (Ryle Telescope), 
%ASM/RXTE count rate during 2005 outburst rise transition of GRS~1915+105.
%Red points A, B and C on the {\it second} panel marked 
%moments at MJD=53416, 53422 and 53442 
%{\it (before,  during and  after radio flares )} respectively.
%{\it Bottom:} PDS of GRS~1915+105 ({\it left column}) plotted versus photon spectrum 
%({\it right column}) for three points A ({\it top}), 
%B ({\it middle}) and C ({\it bottom}) of X-ray light curve.
%}
%\label{radio_pds_spectr}
%\end{figure}

\begin{figure}[ptbptbptb]
\includegraphics[scale=0.8,angle=0]{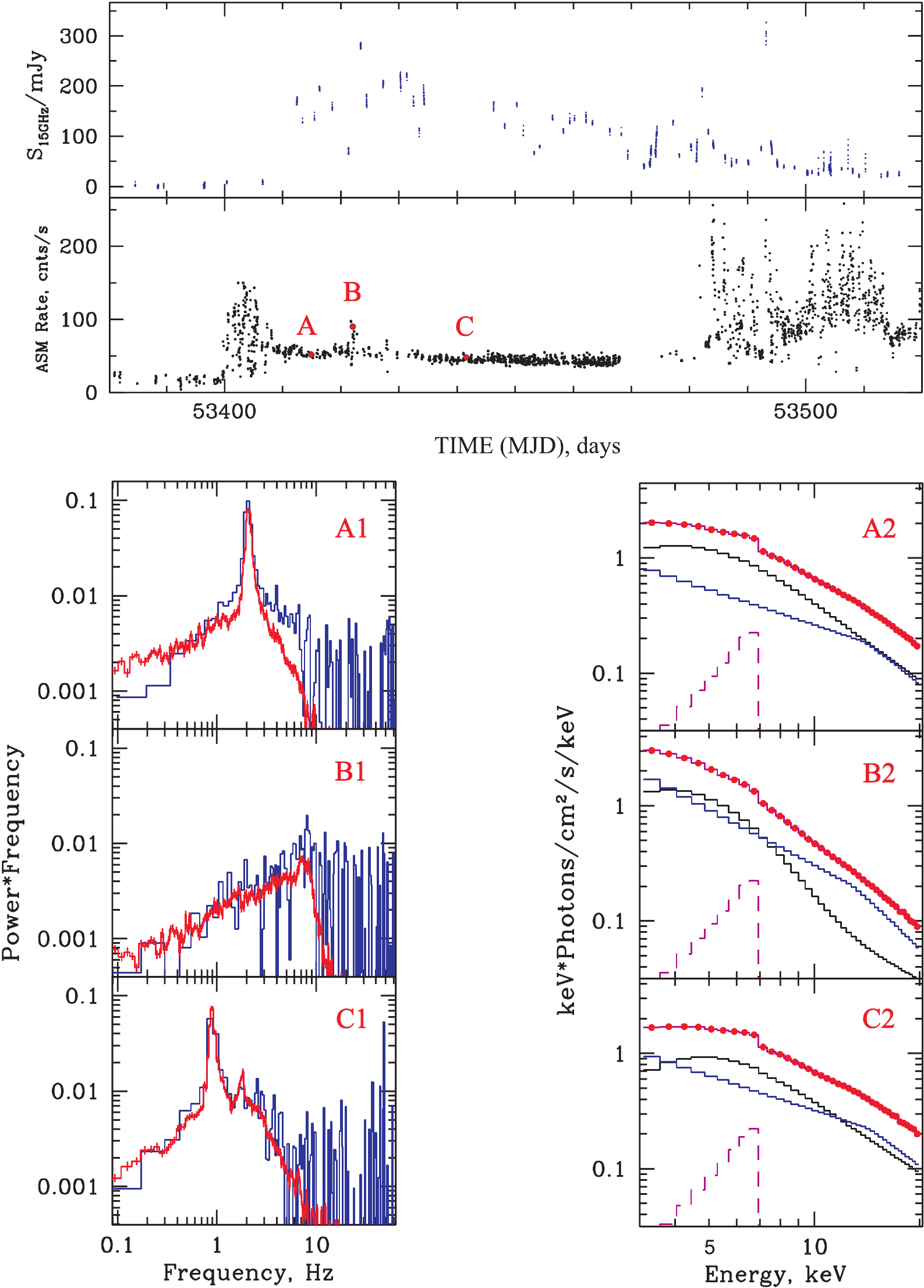}
\caption{{\it Top}: evolution of the flux density $S_{15GHz}$ at 15 GHz (Ryle Telescope), 
\textit{RXTE}/ASM  count rate during  the 2005  rise transition.
{\it Top lower} panel: red points A, B and C    mark 
moments at MJD=53416, 53422 and 53442 {\it (before, during and after radio flares)}
 respectively.
{\it Bottom:} PDSs ({\it left column})  are plotted along with  energy spectral diagram $EF(E)$
({\it right column}) for three points A ({\it top}), 
B ({\it middle}) and C ({\it bottom}) of the X-ray light curve. 
Power density spectra for the soft (red, 3 --15 keV) and hard (blue, 15 -- 30 keV) 
energy bands are presented. There are QPOs at A and C points 
(A1, C1 panels) but there is none at   B point (B1 panel), at the X-ray flare peak. 
 %Fragments of photon spectrum [3-20 keV] demonstrate rising of soft component 
%and Fe K-line complex contribution during radio/X-ray outburst. 
Here data are denoted by red points,
the spectral  model presented with components are shown by 
 blue, black    and 
 dashed purple  lines for {\it BMC1}, {\it BMC2} and {\it laor} components  respectively.
}
\label{radio_appearances}
\end{figure}

\end{document}